# Near-Infrared reflectance spectroscopy of sublimating salty ice analogues. Implications for icy moons.

R. Cerubini*[1], A. Pommerol[1], Z. Yoldi[2], N. Thomas[1]

1) Physikalisches Institut, University of Bern, Switzerland

2) Physics of Ice, Climate and Earth. Niels Bohr Institute. University of Copenhagen, Denmark

*) Corresponding author, (romain.cerubini@space.unibe.ch)

*Abstract*

The composition of the surface of the Galilean icy moons has been debated since the Galileo mission. Several chemistries have been proposed to explain the composition of the non-icy component of the moon's surfaces, notably, sulphuric acid hydrates and magnesium and sodium sulphates. More recently, magnesium and sodium chlorides have been proposed to explain features observed in ground-based observations. We have considered four salts (NaCl, $Na_2SO_4$, $MgSO_4$ and $MgCl_2$) with various concentrations, to produce salty ice analogues. Granular particles were produced by a flash-freezing procedure. Additionally, compact slabs of salty ices were produced by a slow crystallisation of salty liquid solution. These two methods mimic the end-members (plumes and slow ice block formation) for producing hydrated salty ices on the surface of icy moons such as Europa and Ganymede. We have monitored the near-infrared (NIR) evolution of our salty ices during sublimation, revealing differences between the granular particles and the slabs. The slabs formed a higher amount of hydrates and the most highly hydrated compounds. Granular ices must be formed from a more concentrated salty solution to increase the amount of hydrates within the ice particles. The sublimation of salty ices removed all excess water ice efficiently, but the dehydration of the salts was not observed. The final spectra of the slabs were most flattened around 1.5 and 2.0 µm, especially for the $Na_2SO_4$, $MgCl_2$ and $MgSO_4$, suggesting the presence of stable, highly hydrated compounds. We find that $Na_2SO_4$, $MgCl_2$ and $MgSO_4$ are most compatible with the non-icy component at the surface of the icy moons as observed by the NIMS instrument on Galileo and by ground-based observations.

# 1. Introduction

The icy moons of the giant planets show diverse surface features and photometric properties, which suggest that a variety of endogenic and exogenic processes are at play. Of particular interest for our understanding of formation and evolution, are surface properties that can provide clues on moon interiors. The interiors of the Galilean moons have been studied primarily by analysing gravity field measurements from the Galileo spacecraft to determine their moments of inertia. Ganymede and Europa are fully differentiated bodies (Anderson et al., 1998; Collins and Johnson, 2014; Sotin and Tobie, 2004), suggesting the presence of an iron-rich core surrounded by a silicate mantle. Callisto has a slightly higher moment of inertia, which suggests only a partial differentiation with an interior composed of mixed silicates and ice (Anderson et al., 2001; Nagel et al., 2004). The surfaces of the three satellites are mostly water ice. A distortion of Jupiter's magnetic field near each Galilean moon (Khurana et al., 1998; Kivelson, 2000; Kivelson et al., 2002, 1999) is caused by the presence of a conductive layer under the surfaces of the moons, suggesting the presence of a liquid ocean under the ice. The structure of the layers varies between Europa (that has an icy shell above a liquid ocean), and Ganymede and Callisto, in which the global ocean lies between two water ice layers (Sotin and Tobie, 2004; Stevenson, 1996).

The presence of subsurface oceans makes these moons potentially habitable worlds which could harbour extra-terrestrial life forms (Chyba and Hand, 2005; Chyba and Phillips, 2001; Hand et al., 2009). Sampled plume particles from geysers at the south pole of Enceladus revealed the presence of organic compounds and sodium (Postberg et al., 2011, 2009) that is assumed to be endogenic (Waite, 2006; Waite et al., 2017). Plumes have not been directly observed at Ganymede and Callisto, but analysis of ultraviolet observations of Europa obtained with the Hubble Space Telescope (Roth et al., 2014; Sparks et al., 2016) and the reanalysis of previous in-situ measurements by the Galileo spacecraft (Jia et al., 2018) suggest plume activity at Europa. These dynamical events recorded by Hubble and Galileo would involve a nearly continuous pathway between liquid reservoirs and outer space, generating a rapid cooling of liquid droplets. This "flash-freezing" of boiling water implies that ice particles are ejected into space with most eventually re-deposited at the surface.

Geochemical modelling sheds light on the potential composition of oceans that have, and might still interact with a silicate mantle that has evolved from a chondritic parent body. Several authors (Kargel et al., 2000; Zolotov and Kargel, 2009; Zolotov and Shock, 2001) have proposed that the ocean should be enriched in sulphur, chlorine and sodium from hydrothermalism. They also derived the oxidation conditions within the ocean. Other models suggest magnesium sulphate as a relevant component for the briny ocean (Fanale et al., 1982; Kargel, 1991). (Johnson et al., 2019; Vu et al., 2016) constrained the composition of the oceans by modelling the hydrates produced from a solution of $Mg^{2+}$, $Na^+$, $Cl^-$ and $SO_4^{2-}$ with pH, temperature, and ion concentration dependency. They showed a link between the pH and the concentration of Mg ions in the ocean and discussed the endogenic and exogenic source of Mg.

Extensive work has been performed by (Carlson et al., 1996; Carlson, 1999; Moore et al., 2007) on the presence of sulphuric acid hydrates and about the sulphur cycle in general at Europa's surface. They propose a scenario in which the icy shell is enriched in sulphates that are distributed through the proposed geological processes which form disrupted terrains (Collins and Nimmo, 2009; Head and Pappalardo, 1999; Pappalardo et al., 1999; Prockter and Patterson, 2009). The irradiation with iogenic sulphur ions and other energetic particles, leads to the formation of hydrates. The models of reflectance that use optical constants of sulphuric acid hydrates are in good agreement with the NIR spectra of Europa's trailing side (Carlson et al., 2009, 2005).

The hypothesis of the presence of $MgSO_4$ at the surface was formulated and experimentally investigated by (Dalton, 2003; Dalton et al., 2005; McCord et al., 2001, 1999, 1998). The comparison of the reflectance of the non-icy dark component of Europa with laboratory spectra of hydrates was supported by numerical modelling of reflectance data (Dalton et al., 2013; Dalton and Pitman, 2012; Shirley et al., 2010) and by the reanalysis of Galileo spectra (McCord et al., 2010).

The necessity to mix different salts to match the spectra from Europa was suggested, leading to new laboratory efforts to produced flash-frozen solutions of the previously mentioned salts (McCord et al., 2002; Orlando et al., 2005). These results were later used by (Spencer et al., 2006) to match telescopic observations of Europa. More recently, magnesium chloride has been suggested as a precursor to magnesium sulphate at the surface of Europa (Brown and Hand, 2013). Experimental work has shown chloride hydrates at cryogenic temperatures (Hanley et al., 2014) to be relevant for the icy moons.

(Hand and Carlson, 2015; Poston et al., 2017) irradiated NaCl grains with energetic electrons, obtaining a yellowish colouration similar to the one identified on the trailing side of Europa. Recent observations have also reinforced the presence of chlorides at the surface (Fischer et al., 2015; Ligier et al., 2016) as well as the necessity to use salty hydrates to match Ganymede's surface spectra (Ligier et al., 2019). The presence of irradiated NaCl has been again suggested to explain spectral features at the surface of the leading side of Europa (Trumbo et al., 2019).

The nature of Europa's surface was the focus of numerous scientific studies (listed in (Dalton et al., 2010),table 3), with the conclusion that several scenarios were possible, and that the chemistry at the surface was complex. The surface's composition results from (i) endogenous sources of material, (ii) exogenous sources of material (Zahnle et al., 2008), and (iii) energetic particle irradiations of the surface in Jupiter's magnetosphere. Moreover, in addition to the ubiquitous water ice present at the surface, the nature of the dark material is not fully determined; brines of salts have yielded good results in matching the NIR spectra. The deposition of such material on the young and active surface of Europa (Bierhaus et al., 2009; Zahnle et al., 2003) can be mimicked in the laboratory by a flash-freezing procedure (McCord et al., 2002; Orlando et al., 2005). Slow crystallisation processes following thermodynamic equilibrium either at the interface ocean – icy shell, or within the sub-surface hydrogeological systems (Le Gall et al., 2017), are also a relevant procedure to renew the icy surface.

High-quality laboratory spectra of the pure materials form a basis for identification. Optical constants derived from these spectra are the fundamental properties required for the physical modelling of reflectance spectra and are used for quantitative interpretation through modelling of reflectance spectra (Carlson et al., 2009; Dalton, 2007; Dalton et al., 2010, 2013; Dalton and Pitman, 2012; Fischer et al., 2015; Ligier et al., 2016, 2017, 2019; McCord et al., 2010; Shirley et al., 2010). Different key parameters (grain sizes, chemistry, temperature, etc.) affect the optical properties (Grundy and Schmitt, 1998; Schmitt et al., 1998) and mixing materials ultimately results in complex absorption spectra. The photometric properties of mixtures are highly dependent on the way that samples are produced and evolve. Reflectance spectra of complex surfaces, produced using experimental protocols that mimic actual processes expected on planetary surfaces, are therefore desirable to complement laboratory studies of pure compounds and to support reflectance modelling work.

We focus this new study on the production of salty ices and the analysis of their spectral reflectance as they evolve through sublimation. To produce our samples, we use four salts (NaCl, $Na_2SO_4$, $MgSO_4$ and $MgCl_2$) and two different textures: grains and compact slabs which differ in their production by the cooling rate of the solution (slow crystallisation at thermodynamic equilibrium for the slabs versus a "flash-freezing" process for producing granular ices).

Both slabs and granular ices have similar specific absorption features, but these are affected by their structures and surface textures. Slabs represent thick and compact analogues that absorb considerably more light than granular ice. In addition, the high surface-to-volume ratio of small particles makes them prone to faster sublimation and sintering. The presence of salts and the kinetics of crystallisation also affect the structure of the analogues, and ultimately their spectral evolution through sublimation. Studying these two types of analogues allows us to mimic the evolution of end-members of types of ices. The hyperspectral recording of the evolution of the icy, salty analogues that we describe here provides new data to further investigate the surface composition of icy moons, as well as surface or sub-surface processes that drive pristine material production.

# 2. Methods

## 2.1. Samples

In this study, we have investigated four different salt chemistries relevant for the surface of icy moons: NaCl, $MgSO_4$, $MgCl_2$ and $Na_2SO_4$. With each of these salts, we have prepared two types of icy samples: granular ice particles with a mean particle size of ~70µm produced by flash freezing of brines and compact slabs produced by the slow solidification of brines. For each salt and both types of samples, three different concentrations were used (Table 1). The samples produced are regrouped in Figure 1. One should notice the difference of surface texture between the slabs and the granular ices, as well as the overall increase in brightness of the samples after sublimation experiments.

*Table 1: List of chemicals, their sources, types of analogues, and concentrations used.*

| Salt | NaCl | $MgSO_4$ | $MgCl_2$ | $Na_2SO_4$ |
|---|---|---|---|---|
| **Provider (purity)** | Carl Roth© (99%+) | Sigma-Aldrich© (99.5%) | Sigma-Aldrich© (99.5%) | Sigma-Aldrich© (99.5%) |
| **Low concentration brine** | 10wt% Slab, Granular | 10wt% Slab, Granular | 12,5wt% Slab, Granular | 3wt% Slab, Granular |
| **Medium concentration brine** | 20wt% Slab; Granular | 20wt% Slab, Granular | 25wt% Slab, Granular | 15wt% Slab, Granular |
| **Saturated brine** | 30wt% Slab, Granular | 30wt% slab, Granular | 33wt% Slab, Granular | 30wt% Slab, Granular |

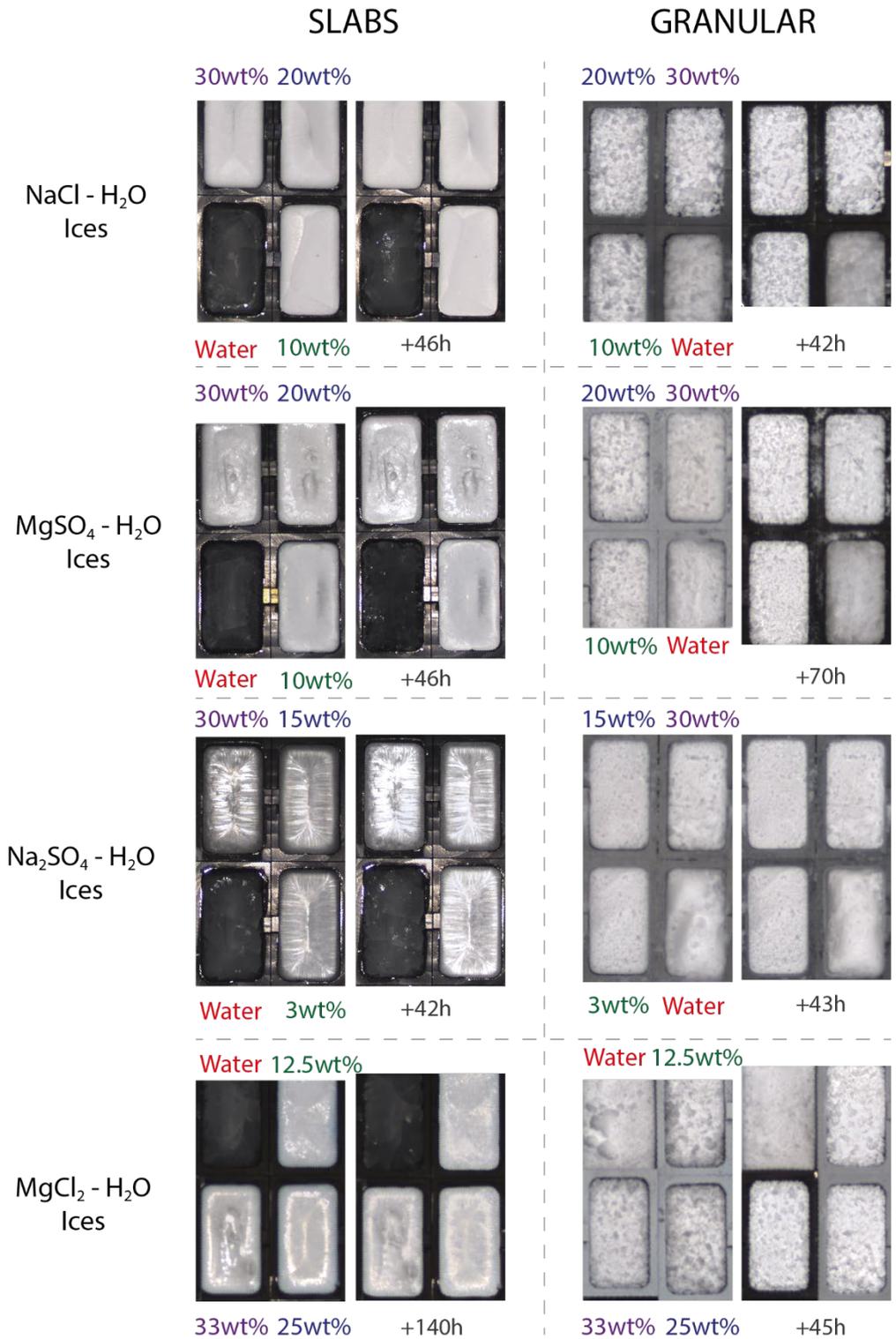

*Figure 1: RGB Images of analogues of salty ices before and after sublimation. The coloured numbers indicate the salt concentration of each sample. The fourth image in the granular sample is of a granular pure water ice. For the slabs and granular ices, a single experimental run was performed for each salt with all three concentrations and a pure water ice sample. Each panel shows the before and after cases for each run. The position of the samples was not changed between the beginning and the end of the experiments.*

### 2.1.1. Granular ice

Granular icy samples have been produced with the Setup for Production of Icy Planetary Analogues (SPIPA) which was fully described by (Pommerol et al., 2019). The general objective of these setups is to provide reproducible and well-characterised icy particles, which can then be used for complementary studies, progressively building a catalogue of their main properties. The setups are portable so that the fresh ice samples can be produced close to the instrument where they will be analysed. For this study, we have used the setup SPIPA-B with an ultrasonic nebulizer that generates particles that attain sphericity (due to surface tension) with diameters of 70 ± 30 µm (Jost et al., 2016, 2013; Yoldi et al., 2015). This size resembles that found at the surface of Europa or Ganymede (Dalton et al., 2012; Ligier et al., 2019; Poch et al., 2018).

Sending droplets of solution directly inside liquid nitrogen is a procedure that generates the so-called Leidenfrost effect, well described in (Feng et al., 2018). In our case, the solution itself is at room temperature (around 293K) as it reaches the vicinity of liquid nitrogen (77K). The temperature difference induces the evaporation of liquid nitrogen. Therefore, the outgassing keeps the particles levitating above the liquid nitrogen for a few seconds. When the temperature difference is not sufficient for the outgassing to maintain the particles levitating, they start to sink inside liquid nitrogen. It then requires a few more seconds for the ice particles to reach the temperature of liquid nitrogen. The estimated cooling rate with the SPIPA-B setup ranges from tens to hundreds of kelvins per second (10 - 100K/s). The particle size also plays a key role; the smaller the particle, the higher the cooling rate. Even here, it does not exceed $10^4$ K/s, this cooling rate is not sufficient to produce amorphous water ice from pure water (Hudait and Molinero, 2014; Ludl et al., 2017; Mayer, 1985).

The concentrations investigated here (Table 1) have been selected following the phase diagram of each salt-water solution (Li et al., 2016; McCarthy et al., 2007). Magnesium-bearing salts usually form several stable hydrated phases (MgSO$_4$·nH$_2$O, with n = 1 to 11). The sodium-bearing salts, on the other hand, usually form only one stable hydrate (NaCl·2H$_2$O). In addition to the three concentrations produced for each salt, a pure water sample was produced as control sample.

With the SPIPA-B setup, a liquid solution of salt and distilled water in the desired concentration is forced through an ultrasonic nebuliser into a bowl with liquid nitrogen (see Fig. 3 in (Pommerol et al., 2019)). This "flash-freezing process" produces spherical ice particles that float for a few seconds at the surface of the liquid nitrogen before sinking and accumulating at the bottom of the bowl. Due to the constant outgassing of liquid nitrogen, water vapor never condenses on the solidifying grains. Furthermore, it takes only a few seconds for the grains to sink into liquid nitrogen and be protected from vapor condensation by the bath of liquid nitrogen.

Once production ends, we cover and place the bowl in a laboratory freezer (at 230K) and wait until the liquid nitrogen has evaporated from the bowl. We then collect the particles with a spoon and sieve (400 microns) them into a rectangular sample holder of dimensions 2x4x2cm to remove the largest aggregates. Finally, the surface is levelled to the rim of the sample holder by scraping it with a spatula to flatten it without compressing it.

During the preparation procedure, all tools and sample holders used are regularly plunged into liquid nitrogen to ensure that they do not warm up and risk metamorphosing the sample upon contact. Performing all the preparations step at the bottom of the cold freezer reduces the contamination of the samples by frost condensed from the ambient air. To prevent any undesired evolution of granular ices, we place them into the vacuum chamber right after finishing the preparation of the sample at the bottom of the freezer.

### 2.1.2. Slabs

Compact slabs of water ice with salts were produced by the slow cooling and crystallisation of a liquid solution at 230K inside a freezer. For each sample, we first prepared 15 ml of liquid solution with the desired concentration. For a given salt, the three concentrations were produced simultaneously to maintain the same conditions of cooling and crystallisation. The three solutions were prepared at ambient lab conditions (~293K), then immediately poured into 2x4x2cm sample holders at ambient temperature. Additionally, a pure water sample was added to act as a control sample. The four sample holders were then installed in a container covered with a lid and placed in the freezer (at 230K) where they remained overnight.

Upon slow crystallisation, the water ice expands and expels the salts from its lattice. Hydrated salt crystals grow at the surface of water ice and within cracks. As a result, all slab samples show similar morphologies and textures with a central "mound" in the middle of the sample holder and a radial organisation of the visible salt crystals around this central mound.

## 2.2. Sublimation experiments

### 2.2.1. Simulation chamber

All sublimation experiments reported here took place in the SCITEAS-2 (Simulation Chamber for Imaging the Temporal Evolution of Analogue Sample) setup (Figure 2); an upgraded version of the SCITEAS setup used in previous sublimation experiments (Jost et al., 2017; Poch et al., 2016; Pommerol et al., 2015). The new setup has many improvements, but the essential modification in this context is the addition of a cold copper plate below the sample, which is cooled by a helium cryocooler (Figure 2, B) down to temperatures as low as 40 K and can also be regulated to higher temperatures if desired. As the original SCITEAS setup, the new chamber has a large quartz window on its upper lid (Figure 2, A) through which characterisation in the visible and infrared spectral ranges can be carried out. The samples are inserted inside the chamber by removing the lid that holds this window. Once the sample is inserted and the lid closed, the chamber is evacuated down to a pressure of $10^{-7}$ mbar using a turbo-molecular pump. The chamber is equipped with pressure and temperature sensors to monitor these parameters continuously.

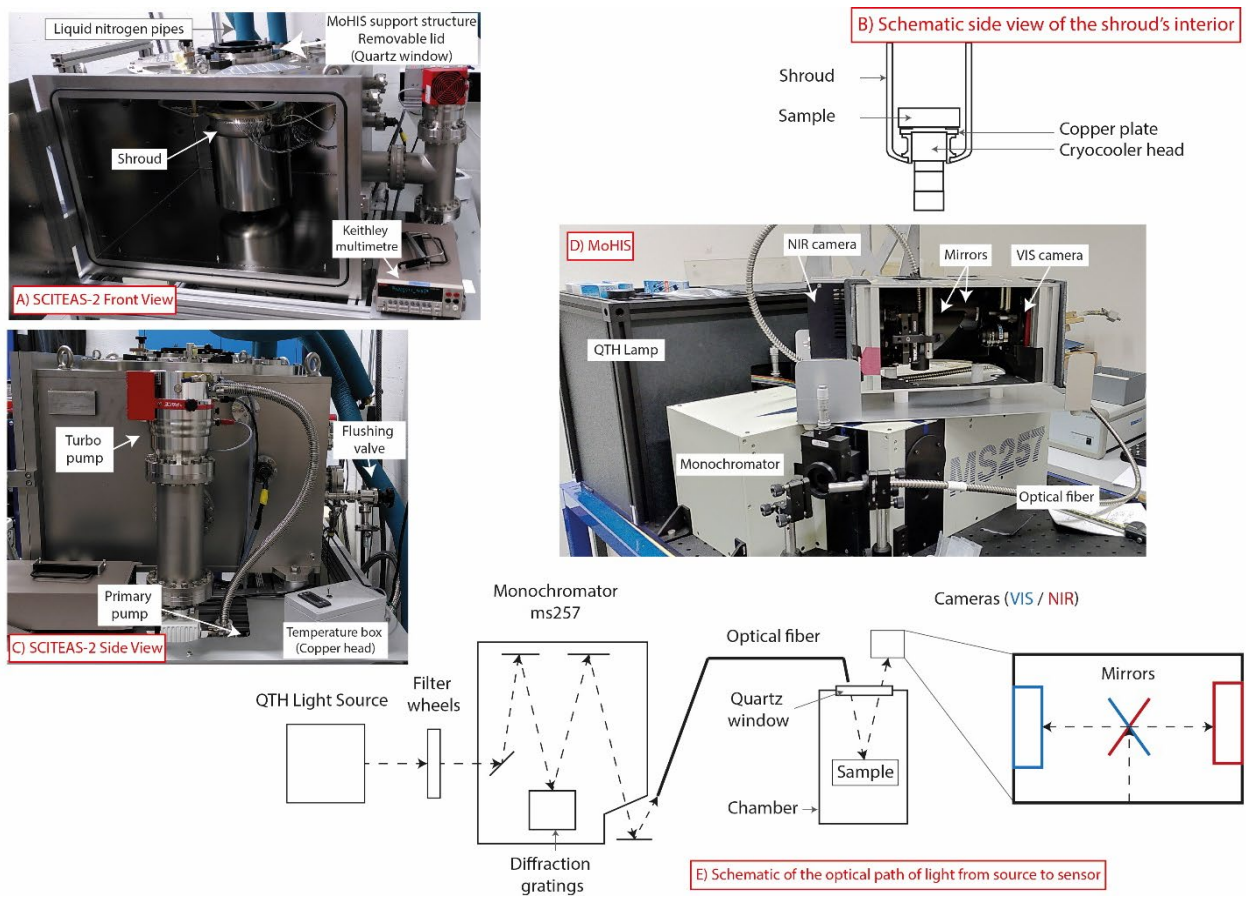

*Figure 2: SCITEAS-2 and the Mobile Hyperspectral Imaging System (MoHIS). A) Front view of the vacuum chamber. The Keithley multimeter reads the tension from the Pt1000 sensors to monitor the temperature. The pressure is measured by a Pfeiffer vacuum© sensor. B) Scheme of a transversal cut inside the shroud. C) Side view of the vacuum chamber with the primary (membrane) and secondary (Turbo-molecular) pumps. D) The hyperspectral imaging system in stand-by configuration. E) Optical path of light from the source to the VIS and NIR cameras.*

### 2.2.2. Hyperspectral imaging setup

The chamber can be opened from the top via a removable cover equipped with a quartz window at its centre (Figure 2, A). Mounted on this window, the Mobile Hyperspectral Imaging System (MoHIS: Figure 2, D) consists of a monochromatic light source that illuminates the sample at selectable wavelengths and two cameras to obtain the spectral image of the reflected radiation (visible or near-infrared). On a trolley next to the chamber, a monochromator (Oriel MS257, Newport©) is coupled with a light source (QTH Source, 50 – 250 W, F2/2 FSA, Newport©) to select different wavelengths. The monochromator has selectable gratings and filters so that the entire available spectral range can be scanned automatically. The visible range is sampled by steps of 15 nm and the near-infrared by steps of 6 nm. The spectral resolution (*full-width-half-maximum*, FWHM of the bandpasses) is either 6.5 nm (visible and near-infrared up to 1700nm) or 13 nm (at wavelengths longer than 1700nm). This latter parameter is adaptable by changing the opening of the entrance and exit slits of the monochromator, which also affects the signal to noise (S/N) ratio. Reducing the opening of the slits will reduce the FWHM as well as it will diminish the exiting light flux, and therefore the light flux to the sample. The light is sent to the surface of the sample via an optical fibre bundle (CeramOptec©) fixed on the structure holding the cameras. The two cameras (a visible CCD camera, model 1501M from Thorlabs and a near-infrared MCT camera, model Xeva2.5 from Xenics) are fixed on a support structure, which can observe the surface of the sample inside the chamber through mirrors.

This entire mesaurement system is controlled by python scripts, which permit high flexibility in terms of the acquisition. Hyperspectral cubes (continuous coverage of the whole selected spectral range with overlap between selected wavelengths) and multispectral cubes (specific selected wavelengths) can be measured. Acquisitions are also occasionally restricted to one single spectral range (VIS or NIR) to better resolve a given absorption feature with increased temporal resolution.

Dark images are acquired during the cube acquisition by closing the monochromator's shutter and are subtracted from the sample images that are averaged at a specific wavelength during the cube acquisition. They are then divided by images of a reference surface made of Spectralon's (Labsphere) at the same wavelength (flat-field correction) to obtain the final images of the sample at a given wavelength. A program allows selection of regions of interest (ROIs) to compute the reflectance value from the pixels contained in the ROI. The final spectrum is a composite of VIS and NIR spectra that we spliced together between 900 and 940 nm.

### 2.2.3. Spectral criteria computation

The normalised integrated band area (NIBA) is computed by integrating the difference between the continuum and the measured values of reflectance, divided by the reflectance of the continuum for each wavelength. The integration is performed using a five-point Newton-Cotes integration formula. We define the continuum as the linear approximation between the two shoulders of the absorption band. This formulation of the NIBA is similar to that of a band depth (1-$R_{band}$/$R_{continuum}$) but includes the entirety of the absorption rather than being calculated at a single wavelength. The selection of the exact wavelengths used will be detailed in Sections 4.2.1 and 4.2.2 for the slabs and the granular ices, respectively.

The centroid of the band (Eq. 1) is determined as the barycentre of the absorption band and is therefore affected by the geometry of the entire band. The centroid will be equal to the location of the maximum of absorption if the band is symmetric.

$$Centroid = \frac{\sum_{i=0}^{n} \lambda_i (R_c - R_b)_{\lambda_i}}{\sum_{i=0}^{n} (R_c - R_b)_{\lambda_i}} \qquad (Eq. 1)$$

With $R_c$ being the reflectance value of the continuum and $R_b$ the reflectance value of the band. This value is used to depict the evolution of the position of the absorption band during the sublimation, taking into account overall shifts and changes in the geometry of the bands

### 2.2.4. Experimental protocol

While the granular samples are prepared, the simulation chamber itself is continuously pre-cooled with liquid nitrogen. For the slabs, the precooling of the chamber took place the day after the sample production, before we insert the samples. The covered sample holders are moved into the simulation chamber, and the cold lid is removed a few seconds before we fix the flange with the window on top of the chamber. We then close the chamber and start evacuating it. Within 5 minutes, the primary pump lowers the pressure down to $10^{-2}$ mbar. The turbo-molecular pump is activated to reduce the pressure down to $10^{-4}$ / $10^{-5}$ mbar in 1 or 2 minutes, and down to $10^{-6}$ mbar in a few additional minutes. AS soon as the pressure reaches $10^{-1}$ mbar, we start the cryocooler. Between the sample holder and the copper plate of the cryocooler, a sheet of graphite was placed to ensure optimal contact, and therefore an optimal cooling.

A total of 12 salty slabs and 12 salty granular ices have been produced to cover a range of concentrations for each salt. The sublimation experiments were conducted for at least 42 hours and a maximum of 140 hours. An example of the temperature and pressure profile during the sublimation experiment is presented in Figure 12 A. Usually, after 42 hours of sublimation, the spectral evolution of the sample has already stabilised. When possible, the sublimation has been conducted on longer periods to monitor eventual long-term changes.

A full hyperspectral cube takes 26 minutes to acquire. The duration of multispectral acquisitions depends on the number of wavelengths to be measured. For specific scans in the NIR, (e.g. as described in section 4.1), intervals of wavelengths can be selected. During our sublimation experiments, the first hyperspectral cube was acquired while the pressure was decreasing from a few millibars to lower than $10^{-3}$ mbar. The second cube was started with a pressure already at $10^{-4}$ mbar. The chamber then reaches $10^{-6}$ mbar and slowly decreases in pressure for the rest of the experiment.

# 3. Results

## 3.1. Reflectance measurements

### 3.1.1. Dry powders used for icy analogues production

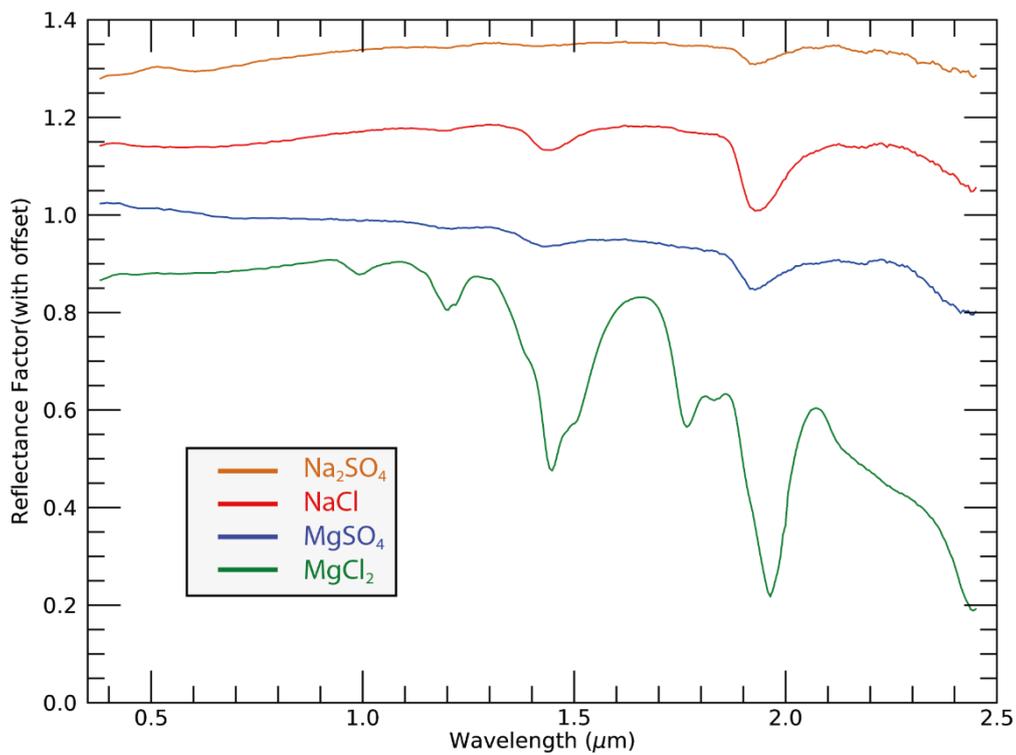

Figure 3 presents the spectra (offset for clarity) of the different salt powders used to prepare the brines in this study. Before the reflectance measurements of the pure chemicals, the powders had spent 24 hours in an oven (333 K) to remove some adsorbed and absorbed water. Still, the presence of water is noticeable in all spectra (bands around 1.4 and 1.9µm) and particularly with $MgCl_2$, which exhibits a spectrum typical of either $MgCl_2 \cdot H_2O$ or $MgCl_2 \cdot 2H_2O$ (by comparison to Hanley et al., 2014, Shi et al., 2019). The three other salts display traces of hydration, probably adsorbed water, which is not part of the crystal lattice of a stable hydrate. In the case of $MgCl_2$, as the exact number of water molecules in the crystal is unknown, all concentrations presented hereafter for the icy analogues are affected by a maximum relative error of 15% corresponding to an absolute error of ±5wt% for a solution at 33wt%. For the other three salts, we estimate the concentrations to be accurate to a level of ±0.5 to ±1 wt%.

The powders of the highly hydrophilic $MgCl_2$ were stored at 350K for a longer period: 15 to 20 hours before their use. The ±5wt% absolute error must be considered as a worst-case for two reasons. Firstly, the salts used for producing the liquid solutions are dissolved in the water a few seconds after being taken out of the oven. This prevents a long interaction with ambient water vapour, as opposed to the exposures of 20 minutes required to measure the spectra. Secondly, the surface of the sample exhibits absorption bands related to water, but this cannot be generalised to the entirety of the thickness of the sample as water is probably more concentrated at the surface. Therefore, we consider reasonable to estimate the absolute error on the concentration of the $MgCl_2$ sample between ±1 or ±2 wt%. The concentrations considered for producing the $MgCl_2$-bearing ices were selected with this bias known. Following the phase diagram of $H_2O$ – $MgCl_2$, a ±2 wt% error on the concentration does not create any overlap between the different cases presented. The chemicals used were systematically dried in an oven with silica beads for at least 15 hours of desiccation at 330K before the solutions were prepared.

### 3.1.2. Analogues: Slabs of salty ices

The reflectance in the visible spectral range (VIS, 0.4 to 0.9 µm) of our sample is not affected by sublimation because both the ice and the salts display a high and constant reflectance in this range. We therefore focus most of the following description on the near-infrared (NIR, 0.9 to 2.4 µm) spectral range.

### 3.1.2.1. Slabs of NaCl

Figure 4 shows the spectra of NaCl – $H_2O$ slabs, slowly crystallised following the thermodynamic equilibrium depicted by the phase diagram of water and NaCl (McCarthy et al., 2007). For all samples, the initial hyperspectral cube recorded shows absorptions of water ice at 1.04, 1.25, 1.5 and 2.0 µm (Schmitt et al., 1998). Absorption features of the dihydrate of NaCl (hydrohalite: $NaCl \cdot 2H_2O$) are also present, before and after sublimation, at around 1.8 µm, 1.98 µm and through the shoulder around 2.2 µm. During the sublimation of these samples, the absorption bands of water ice (1.0, 1.2, 1.5 and 2.0 µm) are significantly reduced. This can be qualitatively assessed by computing the ratio between the initial and final spectra to express absorption features as losses. The similarity between the ratio and the spectra of pure water ice indicates that the evolution of the sample is caused by the sublimation of the water ice. After sublimation, the absorption bands of the *hydrohalite* are more visible as the overlapping absorptions of water ice have vanished. There is however no spectral evidence that the amount of hydrohalite has evolved during the experiments.

Table 2 summarises the positions of the minima of absorption for each spectral signature identified. NaCl salt forms only one known stable hydrate in the conditions investigated (McCarthy et al., 2007): the *hydrohalite* ($NaCl \cdot 2H_2O$). Therefore, the NaCl concentration at the beginning of the experiments does not significantly impact the resulting sublimated sample. It leads to the same hydrate: a salt with low amounts of structural water and sharp absorption features after sublimation. These features are composed of double bands around 1.5, 1.8 and 2.0 µm which have also been observed in previous work (Thomas et al., 2017). The band at 2.17 µm expressed as a shoulder in the initial spectra and well-defined in the final spectra is, to our knowledge, not described in the literature.

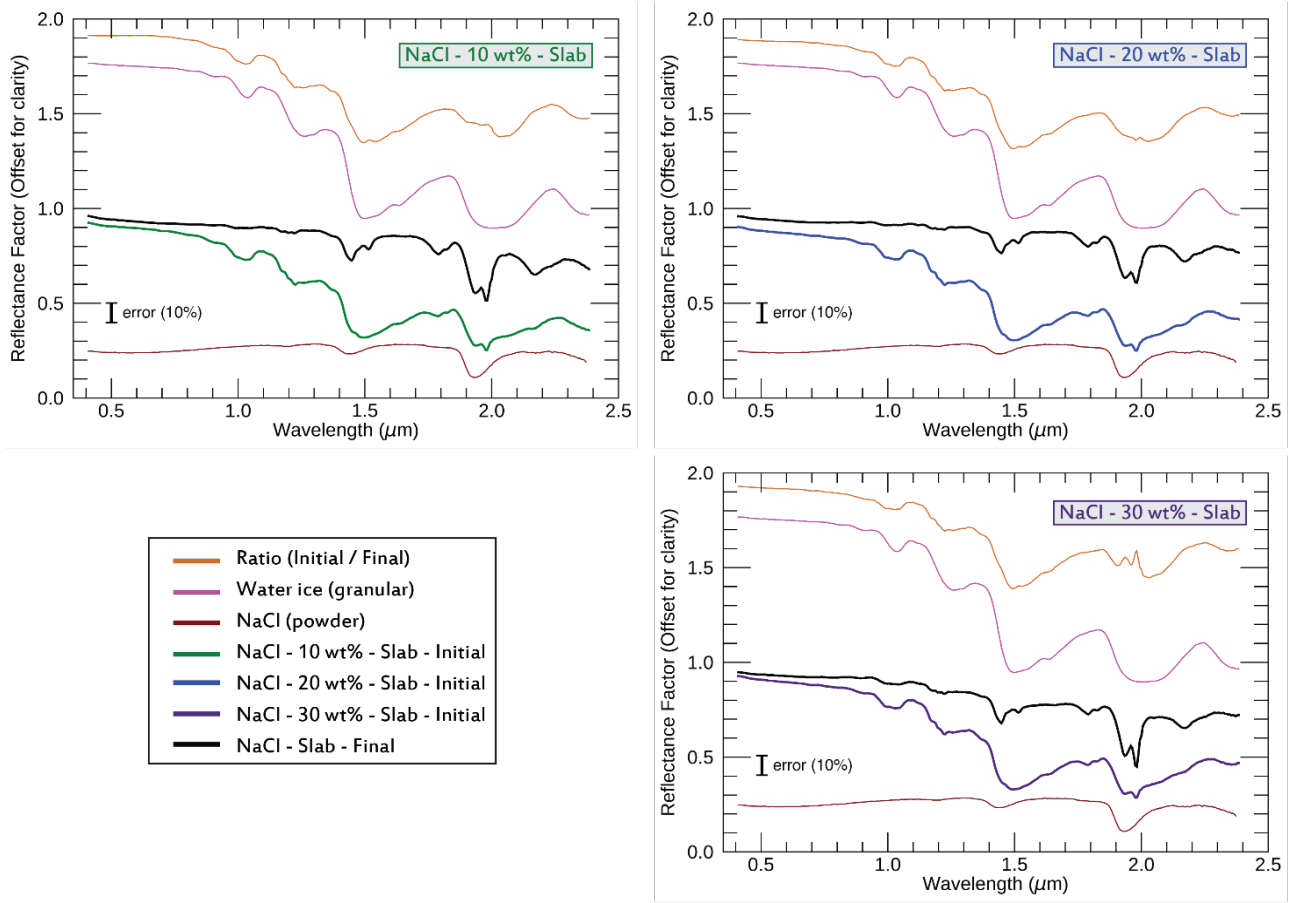

*Figure 4: Reflectance spectra of NaCl slabs before and after 46 hours of sublimation. The top left panel shows the reflectance spectra of the 10 wt% NaCl slab. The top and bottom right panel show the 20 wt% and the 30 wt% NaCl slabs, respectively. The green, blue and purple curves represent the first hyperspectral cubes for each sample; the black curve represents the last hyperspectral cubes acquired 46 hours later. The ratio of initial to final spectra indicates the amount and type of material lost by sublimation. The spectra of water, pure salt and of the ratio have been offset vertically for clarity (+0.85, -0.6, +0.95 respectively).*

Table 2: Positions of the minima of the absorption bands related to water ice and hydrates of NaCl in icy analogues (Figure 4). The values with asterisks are to be considered with caution due to the irregular geometry of the absorption bands.

| Salt | NaCl | | | | | |
|---|---|---|---|---|---|---|
| Concentration | *10 wt%* | | *20 wt%* | | *30 wt%* | |
| Band position for water ice (µm) | Initial Slab | Final Slab | Initial Slab | Final Slab | Initial Slab | Final Slab |
| 0,800 | 0.800* | - | - | - | 0,800 | - |
| 0,920 | 0,905 | 0.905* | 0.905* | - | 0,905 | 0,890 |
| - | - | - | - | - | 0,998 | 0,998 |
| 1,040 | 1,028 | 1.040* | 1,028 | - | 1,028 | 1,040 |
| - | - | - | - | 1,118 | 1,124 | 1,124 |
| - | 1,172 | 1,178 | 1,172 | 1,178 | 1,172 | 1,178 |
| - | 1,196 | 1,196 | 1,196 | 1,196 | 1,196 | 1,196 |
| - | 1,226 | 1,220 | 1,226 | 1,220 | 1,226 | 1,226 |
| 1,262 | 1,256 | - | 1,256 | - | 1,262 | - |
| - | 1.304* | - | 1,304 | - | 1,304 | - |
| - | 1,370 | 1,370 | 1,370 | 1,370 | 1,370 | 1,370 |
| - | 1,454 | 1,448 | 1,454 | 1,448 | 1,454 | 1,448 |
| - | - | - | - | 1,472 | - | 1,478 |
| 1,502 | 1,496 | 1,514 | 1,496 | 1,514 | 1,496 | 1,514 |
| - | 1,646 | - | 1,640 | - | 1,640 | - |
| - | 1,790 | 1,790 | 1,790 | 1,790 | 1,790 | 1,790 |
| - | 1,820 | 1,820 | 1,820 | 1,820 | 1,820 | 1,820 |
| - | 1,934 | 1,934 | 1,940 | 1,934 | 1,934 | 1,934 |
| 2,000 | 1,982 | 1,982 | 1,982 | 1,982 | 1,976 | 1,982 |
| - | 2,156 | 2,174 | 2,156 | 2,174 | 2,174 | 2,168 |

### 3.1.2.2. Slabs of MgCl$_2$

Figure 5 presents the reflectance spectra of the slabs prepared with MgCl$_2$ at three concentrations. The concentrations have been selected following the phase diagram of MgCl$_2$ and water (Davis et al., 2009; Li et al., 2016). All initial spectra present shallow absorption bands related to water ice at 1.04, 1.25 and 1.5 µm. The stronger absorption band of water ice at 2.0 µm can also be recognised in all samples despite its strong distortion because of the presence of hydrates. The band at 1.65 µm cannot be identified. The positions (minima) of the bands are provided in Table 3.

After 140 hours of sublimation, we witness a flattening of the 1.5 µm bands and an overall increase of the reflectance. The ratio of the initial over the final spectra, which depicts losses of water as absorptions bands on this spectrum, shows absorption features of water at 1.04 and 1.25 µm. Around 2.0 µm, a band has grown for every sample, producing a double absorption band in this region of the spectrum. By comparison with the spectrum of the initial MgCl$_2$ salt (with a low degree of hydration), the band at 1.96 µm is superimposed onto one of the hydrates of MgCl$_2$; the shoulder at 2.01 µm is due to the presence of water ice. The absorption complex at 1.5 µm is not strictly flat in every sample, but the absorption features are not sufficiently pronounced to be identified.

For concentrations of 12.5 and 25 wt%, only the highly-hydrated phase MgCl$_2$·12H$_2$O was expected to form. Around saturation, which is the case for the 33wt% solution, the octahydrate MgCl$_2$·8H$_2$O is also expected to form. These two highly hydrated forms of MgCl$_2$ present, when not mixed with ice, the absorption feature at 1.96 µm identified in the slabs. They also exhibit a simple or double absorption feature around 1.77 µm in addition to the absorption features related to water (roughly at 1.00, 1.20 and 1.43 µm) (Shi et al., 2019).

Even following a long sublimation experiment, the spectral features remain relatively flat, indicating that the sublimation's conditions were not sufficient to alter significantly the hydrates formed.

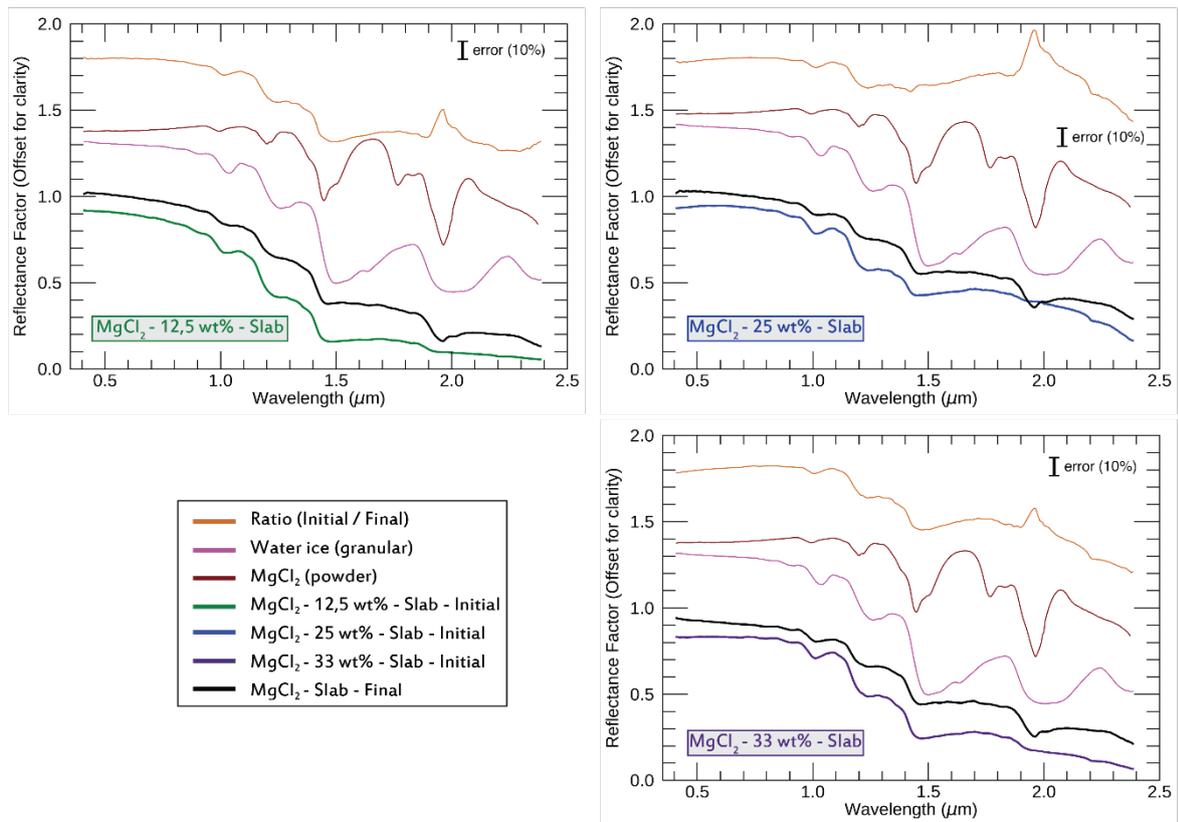

*Figure 5: Reflectance spectra of MgCl$_2$ slabs before and after 140 hours of sublimation. The top left panel shows the reflectance spectra of the 12.5 wt% MgCl$_2$ slab. The top and bottom right panel show the 25 wt% and the 33 wt% MgCl$_2$ slab, respectively. The green, blue and purple curves represent the first hyperspectral cube for each sample; the black curve represents the last hyperspectral cube acquired 140 hours later. The ratio of initial to final spectra indicates the amount and type of material lost by sublimation. Only the spectra of water, pure salt and the ratio have been offset vertically for clarity (for 12.5 and 33 wt%: water +0.40, pure salt +0.50 and ratio +0.90; for 25wt%: water +0.50, pure salt +0.60, and ratio +0.87).*

Table 3: Positions of the minima of the absorption bands related to water ice and hydrates of MgCl$_2$ in icy analogues (Figure 5). The values with stars are to be considered with caution due to the irregular geometry of the absorption bands.

| Salt | MgCl$_2$ | | | | | |
|---|---|---|---|---|---|---|
| Concentration | *12.5 wt%* | | *25 wt%* | | *33 wt%* | |
| Band position for water ice (µm) | Initial Slab | Final Slab | Initial Slab | Final Slab | Initial Slab | Final Slab |
| 0.800 | - | - | - | - | - | - |
| 0.920 | 0.905 | 0.905 | 0.920* | 0.920* | 0.920 | 0.905 |
| 1.040 | 1.016 | 1.016 | 1.028 | 1.046* | 1.010 | 1.016 |
| 1.262 | 1.238 | 1.238* | 1.256 | 1.256* | 1.238 | 1.244 |
| 1.502 | 1.448 | 1.468 | 1.478 | 1.466 | 1.472 | 1.466 |
| 2.000 | 1.946* | 1.958 | 1.946 | 1.964 | 1.964* | 1.964 |
|  |  | 2.012 |  | 2.012 |  | 2.012 |

### 3.1.2.3. Slabs of MgSO$_4$

The reflectance spectra of icy slabs made from a solution of MgSO$_4$ are shown in Figure 6. For every salt concentration, the first spectrum shows the absorption bands of water at 1.04, 1.25, 1.5 and 2.0 µm. The 1.65 µm band of crystalline water ice is also present, although the water band at 1.5 µm partially hides it. During the sublimation, the trend is the same as for all other samples: the water bands sharpen, evolve towards a V-shape and shift slightly toward shorter wavelengths; the absolute values of reflectance increase. The higher the concentration of MgSO$_4$, the more differentiated the bands appear after sublimation. In particular, the 30 wt% MgSO$_4$ slab exhibits absorption features around 1.5 µm (at 1.49, 1.62 and 1.77 µm, cf. Table 4) and 2.0 µm (1.95, 2.00 and 2.05 µm, see Table 4), as observed for brines of MgSO$_4$ at low temperature by (Dalton et al., 2012, 2005).

The initial-to-final ratio shows the loss of water ice, the loss of two absorption bands located at 1.9 and 2.1 µm, and a small feature between 2.3 and 2.4 µm. The two bands are specific to hydration of sulphates and the feature between 2.3 and 2.4 is most likely related to the sulphate anions (SO$_4^{2-}$) in a hydrated environment (Gendrin et al., 2005).

The hydrates that are expected to form with the concentrations tested are limited to a few candidates. Following the thermodynamic equilibrium previously presented (Hogenboom et al., 1995; McCarthy et al., 2007; Peterson and Wang, 2006), the 10 and 20 wt% $MgSO_4$ solutions would lead to the formation of *meridianiite* ($MgSO_4 \cdot 11H_2O$). The 30 wt% solution would contain, as a suspension before any cooling down, some *epsomite* ($MgSO_4 \cdot 7H_2O$). The slow cooling rate applied for the slab production leads to the presence of both *epsomite* and *meridianiite* within the analogue. The epsomite explains the spectral differences of this last sample compared to those with lower concentrations

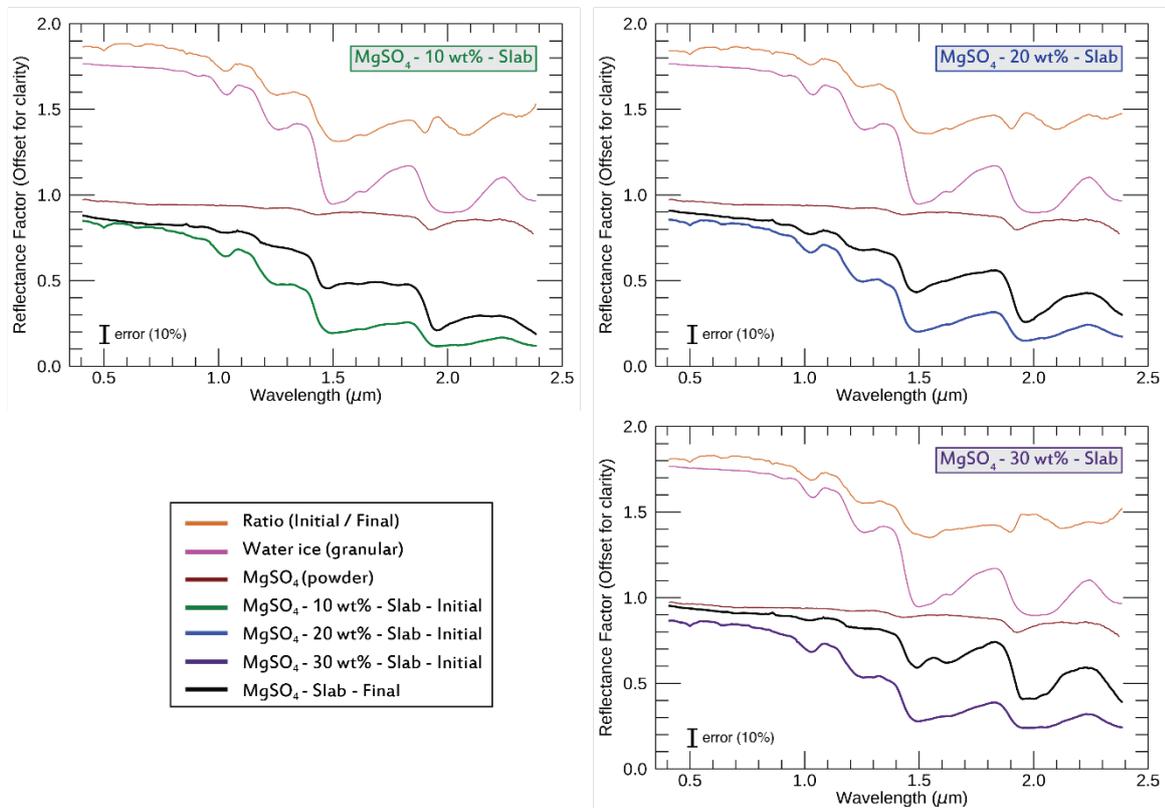

*Figure 6: Reflectance spectra of $MgSO_4$ slabs before and after a 46-hour sublimation experiment. The top left panel shows the reflectance spectra of the 10 wt% $MgSO_4$ slab. The top and bottom-right panels show the 20 wt% and the 30 wt% $MgSO_4$ slabs, respectively. The green, blue and purple curves represent the first hyperspectral cube for each sample; the black curve represents the last hyperspectral cubes acquired 46 hours later. Only the spectra of water, pure salt and the ratio have been offset vertically for clarity (+0.85, -0.05, +0.90 respectively).*

Table 4: Positions of the minima of the absorption bands related to water ice and hydrates of MgSO$_4$ in icy analogues (Figure 6). The values with stars are to be considered with caution due to the irregular geometry of the absorption bands.

| Salt | MgSO$_4$ | | | | | |
|---|---|---|---|---|---|---|
| Concentration | *10 wt%* | | *20 wt%* | | *30 wt%* | |
| Band position for water ice (μm) | Initial Slab | Final Slab | Initial Slab | Final Slab | Initial Slab | Final Slab |
| 0,800 | - | - | - | - | - | - |
| 0,920 | - | - | - | 0.905* | - | - |
| 1,040 | 1,028 | 1.040* | 1,028 | 1,028 | 1,028 | 1,016 |
| 1,262 | 1,256 | - | - | 1,256 | 1,256 | - |
| - | - | - | - | - | 1.304* | - |
| 1,502 | 1,496 | 1,478 | 1,496 | 1,490 | 1,496 | 1,490 |
| - | - | 1,610 | 1,628 | - | 1,628 | 1,622 |
| - | 1,748 | 1,778 | - | - | - | 1,772 |
| 2,000 | 1,952 | 1,952 | 1,952 | 1,964 | 1,976 | 1,952 |
| 2,000 | - | | - | - | - | 2,000 |
| - | - | - | - | - | 2,048 | 2,054 |

### 3.1.2.4. Slabs of $Na_2SO_4$

The reflectance spectra of the slabs of water ice with $Na_2SO_4$ are shown in Figure 7, and the positions of the minima of the absorption bands are given in Table 5.

The initial reflectance spectra of slabs with 3 and 15 wt% of $Na_2SO_4$ exhibit absorption bands related to water at 1.04, 1.25, 1.5 and 2.0 µm although strongly flattened, almost saturated by the presence of the hydrated salt. The oversaturated slab with 30 wt% $Na_2SO_4$ shows an overall slope over the NIR. The bands at 1.04 and 1.25 µm are shallower than those from lower concentrations. The water absorption bands at 1.5 and 2.0 µm are flattened due to the salt and not recognisable.

The 42 hours of sublimation have resulted in two different types of behaviour. For the 15 and 30 wt% $Na_2SO_4$ slabs, the sublimation has produced an overall offset of the reflectance values without major modifications of the spectra. By looking at the initial over the final spectra, the 30 wt% case is noisy and does not exhibit any spectral signature that could be differentiated from noise. The 15 wt% case shows the loss of water, as bands are present at 1.04, 1.25 and 1.5 µm, without leading to noticeable modifications of the spectra over the NIR region.

For the 3 wt% $Na_2SO_4$ slab, the 42 hours of sublimation have led to the appearance of a strong absorption band at 1.94 µm. The ratio indicates the loss of water ice.

From the phase diagram of $Na_2SO_4$ and water (McCarthy et al., 2007, adapted from Kargel, 1991), the 3 wt% case is undersaturated, the 15 wt% case is close to saturation at ambient condition, whereas the 30 wt% case is oversaturated. The number of stable hydrates for $Na_2SO_4$ is limited to *mirabilite* ($Na_2SO_4 \cdot 10H_2O$). This hydrate is very rich in water; therefore, the water absorptions on the initial spectra are distorted. $Na_2SO_4$ is either strongly hydrated or anhydrous, which is one of the main difference compared to the other sulphate tested in this study. In the case of $MgSO_4$, the hydration state can be reduced by the transition from *meridianiite* to *epsomite* (Figure 9; 10, 20 and 30 wt%). In the case of $Na_2SO_4$, the sublimation experiment performed has led to the sublimation of a bit of water ice without strong modifications of the spectra. The 3 wt% case exhibits small features on the ratio at 1.9 and 2.1 µm, perturbed by the growth of the band at 1.94 µm.

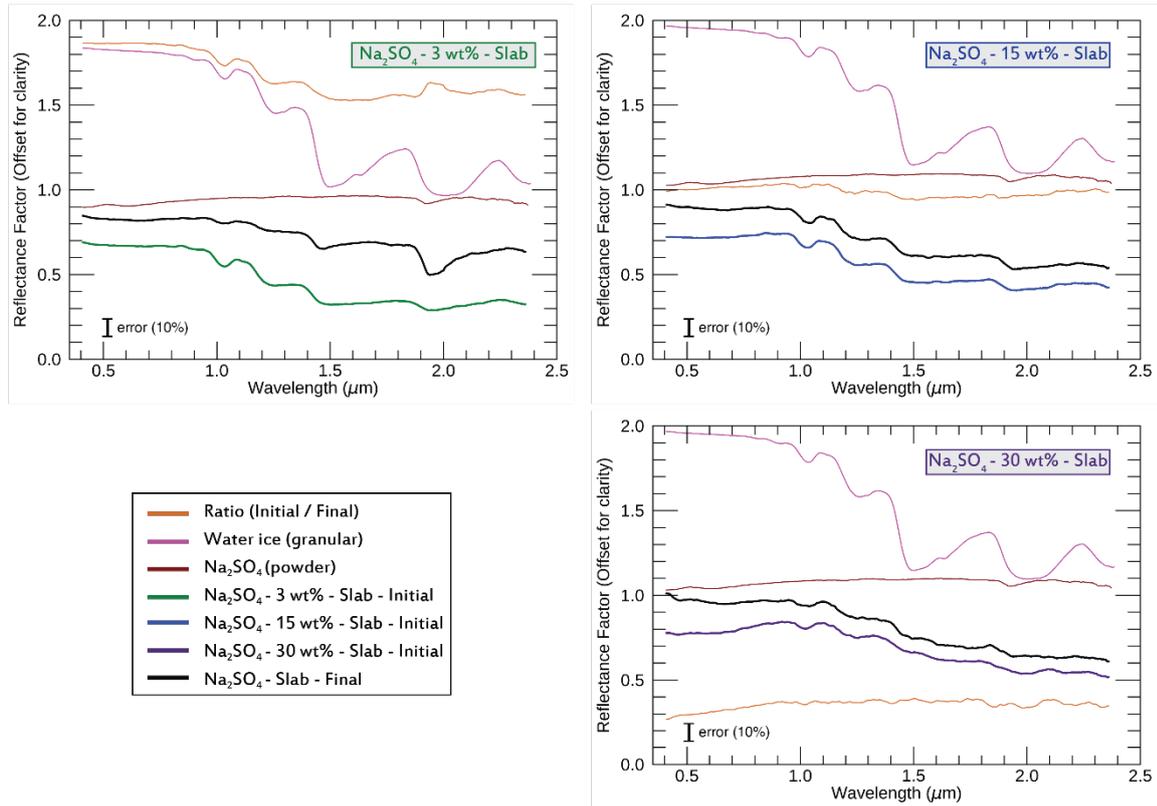

*Figure 7: Reflectance spectra of Na$_2$SO$_4$ slabs before and after a 42-hours sublimation experiment. The top left panel shows the reflectance spectra of the 3 wt% Na$_2$SO$_4$ slab. The top and bottom-right panels show the 15 wt% and the 30 wt% Na$_2$SO$_4$ slabs, respectively. The green, blue and purple curves represent the first hyperspectral cube for each sample; the black curve represents the last hyperspectral cube acquired 42 hours later. The spectra of water, pure salt and the ratio have been offset vertically for clarity (for 3 wt%: +0.92, +0.11, +1.05; for 15 wt%: +1.05, +0.24, +0.20; for 30 wt%: +1.05, +0.245, -0.5).*

Table 5: Positions of the minima of the absorption bands related to water ice and hydrates of Na$_2$SO$_4$ in icy analogues (Figure 7). The values with stars are to be considered with caution due to the irregular geometry of the absorption bands.

| Salt | Na$_2$SO$_4$ | | | | | |
|---|---|---|---|---|---|---|
| Concentration | *3 wt%* | | *15 wt%* | | *30 wt%* | |
| Band position for water ice (µm) | Initial Slab | Final Slab | Initial Slab | Final Slab | Initial Slab | Final Slab |
| 0,800 | 0,815 | - | - | - | - | - |
| 0,920 | 0,905 | 0.890 - 0.905 | 0.890* | 0,890 | - | - |
| 1,040 | 1,034 | 1.029 - 1.052 | 1,034 | 1,046 | 1,022 | 1,040 |
| 1,262 | 1,256 | 1.256* | 1.250 - 1.256* | 1.286* | 1.256* | - |
| 1,502 | 1,502 | 1,472 | 1.490* | 1.490* | - | - |
| 2,000 | 1,934 | 1,941 | 1.952* | 1.952* | - | - |

### 3.1.3. Analogues: Granular salty ices (SPIPA-B)

Similarly to the slab samples investigated in Section (3.1.2), the spectra of the particles produced for this set of experiments do not evolve significantly in the VIS range upon sublimation. The description will therefore be focused on the NIR region.

Figure 8, 9, 10 and 11 present reflectance spectra of the granular salty ices. They all exhibit absorption features of water ice (1.04, 1.25, 1.50, 1.65 and 2.00 µm) in the first spectra. The other specific absorption features related to each type of salt will be discussed in the following sub-sections.

#### 3.1.3.1. Granular ices: NaCl

In addition to the water ice features, spectral signatures diagnostic of the presence of *hydrohalite* are also discernible within these analogues (Figure 8). The absorption features of NaCl increase with the concentration in the mixtures; the sample with 30wt% shows stronger features around 1.8, 1.98 and 2.2 µm than the other samples.

During the sublimation, the reflectance increases, especially in the NIR. The decrease in water-related absorption bands indicates the loss of water by sublimation. The initial-to-final ratio shows features specific of *hydrohalite* around 1.8 µm and at 1.98 µm. This indicates that, together with the loss of water ice, a partial dehydration of *hydrohalite* ($NaCl·2H_2O$) into *halite* (NaCl) occurred. This is consistent with the observed tendency of a general flattening of the spectrum, transitioning from a spectrum dominated by water ice to a spectrum dominated by anhydrous NaCl (Figure 3, red curve; Figure 8, brown curve).

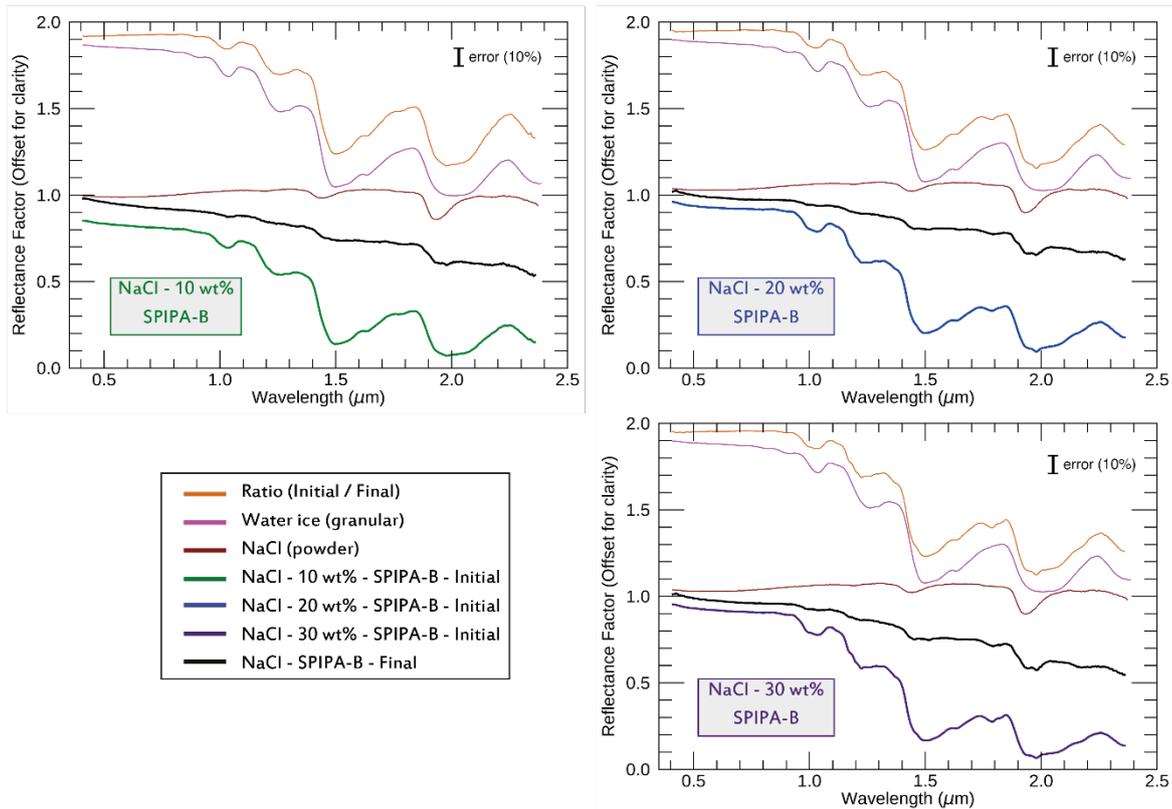

*Figure 8: Reflectance spectra of granular ices of NaCl before and after a 42-hours sublimation experiment. The top left panel shows the reflectance spectra of the 10 wt% NaCl SPIPA-B. The top and bottom right panel show the 20 wt% and the 30 wt% NaCl SPIPA-B, respectively. The green, blue and purple curves represent the first hyperspectral cubes for each sample; the black curve represents the last hyperspectral cubes acquired 42 hours later. The spectra of water, pure salt, and the ratio have been offset vertically for clarity (for 10 wt%: +0.95, +0.15, +1.05; for 20 wt%: +0.98, +0.19, +1.01; for 30 wt%: +0.97, +0.19, +1.00).*

### 3.1.3.2. Granular ices: $MgCl_2$

Absorption features related to $MgCl_2$ are barely noticeable in the initial spectra (Figure 9). When looking at the largest concentration (33 wt%), small features can be identified. Two small ones are located at 1.72 and 1.82 µm (seen in $MgCl_2·12\ H_2O$ in Shi et al., 2019, Fig. 8) as well as a shoulder at 2.2 µm (seen in $MgCl_2·[8, 12]H_2O$). As with NaCl, the higher the concentration, the stronger the hydrate-related absorption characteristics.

Upon sublimation, the reflectance spectra evolved significantly for all salt concentrations, and the absorption bands related to a lower hydration state appeared. The bands at 1.04 and 1.25 µm are reduced in intensity, especially for lower salt concentrations. The absorption complex at 1.5 µm shifts towards 1.45 µm and is sharper with higher amount of salt. The shoulder at 1.69 µm is also more pronounced when the salt concentration is higher. The absorption complex at 2.0 µm transforms into a strong band centred at 1.96 µm with a smaller feature at 2.01 µm. The feature at 2.2 µm disappears.

All these observations allow us to assign the final spectra to lower hydration states of the salt. For the 25 and 33 wt% samples, the presence of the shoulder at 1.69 µm is a strong argument for a hydration state equal or lower to $MgCl_2 \cdot 6H_2O$ (Shi et al., 2019).

The case of the sample prepared with the 12.5 wt% solution is not straightforward, as the 1.69-shoulder is smaller and slightly displaced to 1.71 µm. Two small absorption features are also located in this region, at 1.55 and 1.64 µm. The spectral features observed with this concentration are related to $MgCl_2 \cdot 6H_2O$. Water ice still dominates the 1.5-µm absorption complex. The spectral ratios show evidence for the loss of water and the growth of previously discussed features at around 2.0 µm.

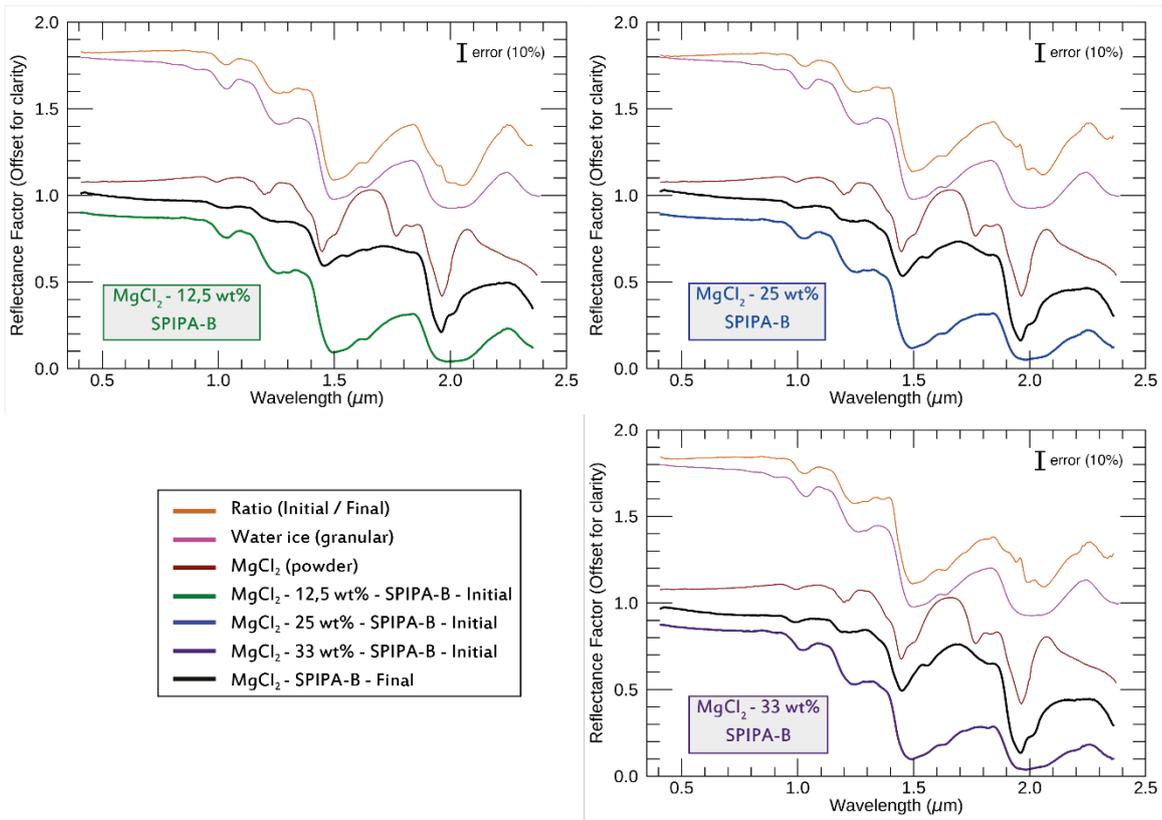

*Figure 9: Reflectance spectra of the granular ice made from a liquid solution of water with 12.5 (green), 25.0 (blue) and 33.0 wt% (purple) $MgCl_2$. The green (blue, and purple) curves represent the first hyperspectral cube. The black curves represent each time the final hyperspectral cubes for each sample made after 45 hours. The spectra of water, pure salt, and the ratio have been offset vertically for clarity (+0.88, +0.20, +0.94 respectively).*

### 3.1.3.3. Granular ices: MgSO$_4$

Prior to sublimation, the granular ices prepared with MgSO$_4$ (Figure 10) do not show the strong absorption features at 1.9 or 2.1 µm related to the pure hydrates of MgSO$_4$ (Gendrin et al., 2005). The first spectrum is dominated by water ice, and the higher the amount of salt within the solution, the flatter the absorption complexes at 1.5 and 2.0 µm are.

The spectrum of the sample prepared with 30 wt% MgSO$_4$ presents a tiny absorption feature at 1.77 µm, probably related to MgSO$_4$·[2, 3 or 12]H$_2$O (*kieserite* and *sanderite* for 2 and 3 H$_2$O molecules, respectively) by comparison with (Dalton et al., 2005). This feature is weak in the 20 wt% spectrum and absent from the 10 wt% spectrum. After sublimation, the overall reflectance values increase. The absorption bands of water at 1.04 and 1.25 µm are stronger for samples with higher salt concentrations. The spectra for the 10 and 20 wt% samples appear similar, showing a flat 1.5-µm complex and a "V" shaped 2.0-µm absorption complex centred at 1.95 µm. A small shoulder at 2.07 µm, previously observed and attributed to epsomite (Brown and Hand 2013) is barely identifiable. The 30 wt% spectrum still shows the two features at 1.64 and 1.77 µm in its 1.5-µm absorption complex. The 1.64-µm feature could be attributed to either a remain of the 1.65-µm band of crystalline water ice, *hexahydrite* (MgSO$_4$·6H$_2$O) or MgSO$_4$·12H$_2$O, as shown in (Dalton, 2007).

The spectral ratio calculated for all concentrations mimics the spectrum of the pure granular water ice (Figure 10, pink curves), which demonstrates that the sublimation results in the loss of water ice without a change in the hydration state of the salt.

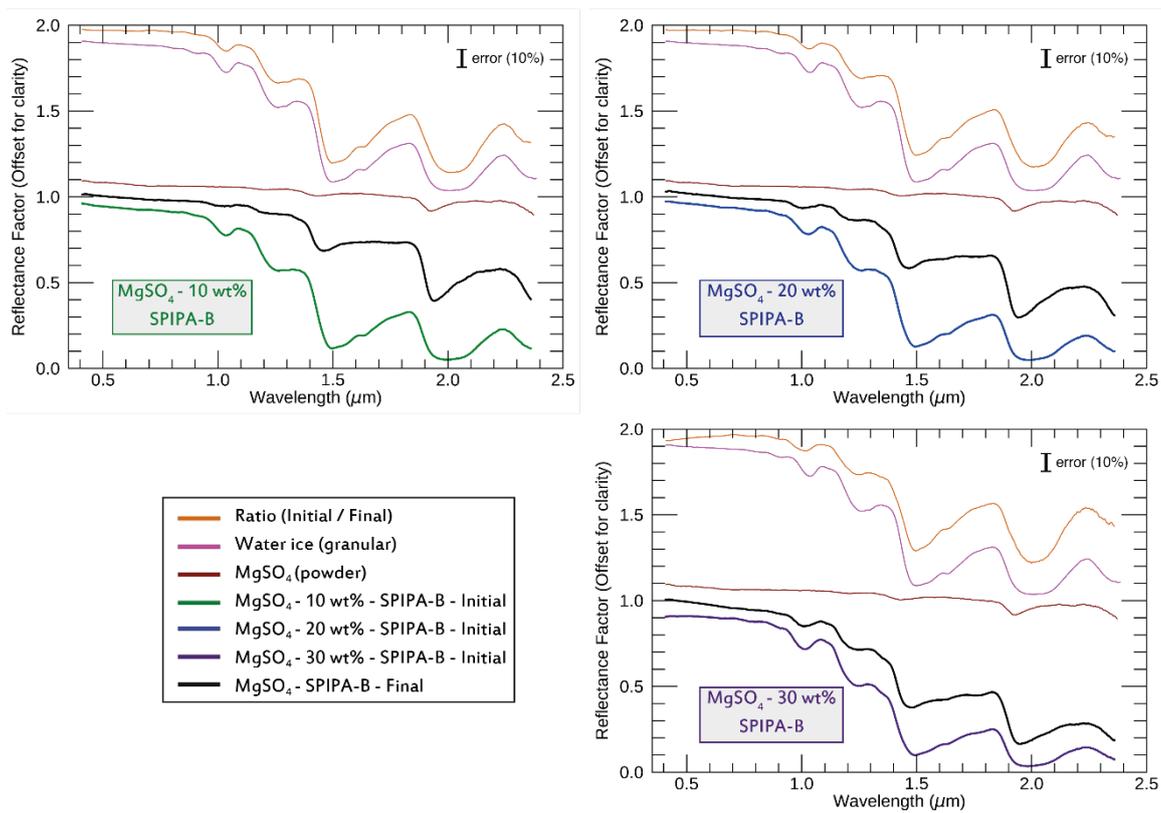

*Figure 10: Reflectance spectra of granular ices with MgSO$_4$ before and after a 70-hours sublimation experiment. The top left panel shows the reflectance spectra of the 10 wt% MgSO$_4$ sample. The top and bottom right panel show the 20 wt% and the 30 wt% MgSO$_4$ samples, respectively. The green, blue and purple curves represent the first hyperspectral cube for each sample; the black curve represents the last hyperspectral cube acquired 70 hours later. The spectra of water, pure salt and the ratio have been offset vertically for clarity (+0.99, +0.07, +1.03 respectively).*

### 3.1.3.4. Granular ices: Na$_2$SO$_4$

The spectra of samples produced with Na$_2$SO$_4$ solutions (Figure 11) are dominated by water ice. Shallow absorption bands related to the hydrates are noticeable, especially for the highest concentration of salt (30 wt%).

On the first 30wt% spectrum, a band is present at 1.77 μm as well as two smaller features at 2.18 and 2.23 μm. These features have been previously assigned to *mirabilite* by (Dalton, 2007; Dalton et al., 2005). They are barely identifiable in the spectra of the 15 wt% salt concentration and lower. During the sublimation sequence, the spectra for all concentrations flatten as the features of water ice vanish. The final spectra for the 15 and 30 wt% concentrations display reduced absorption features of water ice. The absorption band at 1.77 µm is still present but also reduced. By looking at the spectral ratios, the loss of water ice becomes more noticeable in the case of the 30 wt% salt concentration, as well as the loss of the absorption band at 1.77 µm. In the sample with 15 wt% salt concentration, the band at 1.77 um does not change with sublimation.

For the 3 wt% sample, the absorption complexes at 1.5 and 2.0 µm are partially saturated due to the presence of the hydrated salt that broadens the absorption bands of water. Nevertheless, water ice seems to be still dominant, whereas for cases that are more concentrated the presence of the hydrated salt dominates the spectral signature of the analogue by strongly deforming the water ice bands.

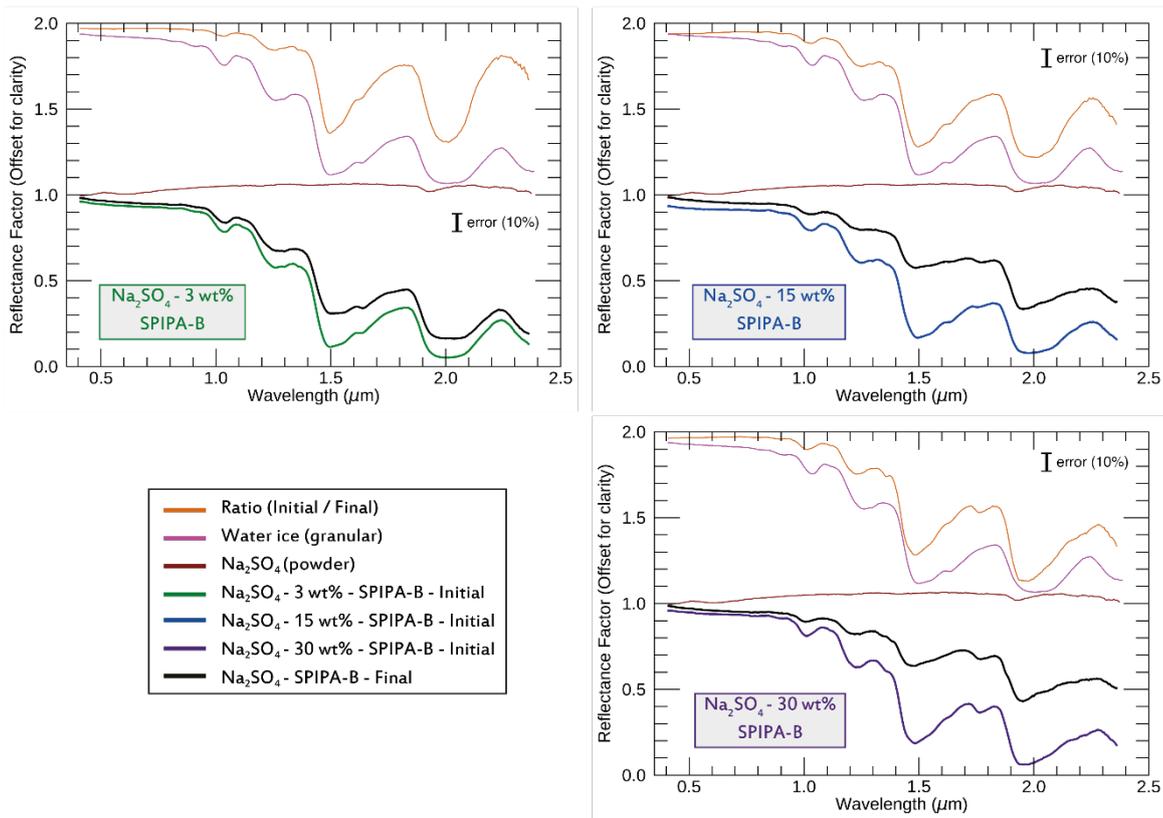

*Figure 11: Reflectance spectra of granular ices of Na₂SO₄ before and after a 43-hour sublimation experiment. The top left panel shows the reflectance spectra of the 3 wt% Na₂SO₄ ice sample. The top and bottom right panel show the 15 wt% and the 30 wt% Na₂SO₄ sample, respectively. The green, blue and purple curves are the first hyperspectral cube for each sample; the black curve is the last hyperspectral cube acquired 43 hours later. The spectra of water, pure salt, and the ratio have been offset vertically for clarity (for 3 wt%: +1.05, +0.24, +0.99; for 15 and 30 wt%: +1.02, +0.21, +0.99).*

# 4. Discussion

## 4.1. Crystallinity of the pure ice and granular salty ice particles

We derive the cooling rate of our samples from the Leidenfrost effect, probably being tens to hundreds kelvin per second (Section 2.1.1), and therefore not sufficient to generate amorphous ice. The bulk of the SPIPA-B ice particles must therefore consist of crystalline $I_h$ water ice. Nevertheless, the first, short contact of the water droplet with liquid nitrogen might induce a higher cooling rate than during the rest of the process. It is therefore possible to foresee a thin shell of amorphous ice surrounding a core of crystalline ice. In addition, water vapour from the atmosphere in the lab could also directly condense at the surface of the liquid nitrogen or on the cold walls of the steel bowl, contaminating the sample with a small fraction of amorphous ice.

We have monitored the NIR reflectance of pure SPIPA-B ice particles to assess the crystalline vs amorphous nature of these particles. We first kept the sample at the lowest possible temperature, and then warmed it up progressively. The aim is to detect a possible evolution of the absorption band at 1.65 µm, a band diagnostic of crystalline water ice but absent in amorphous ice. The band is, however, also affected by temperature, being deeper at low temperature (Grundy and Schmitt, 1998). The initial absence of this band, followed by its appearance and growth would indicate that the water ice produced is initially amorphous and then progressively crystallises.

In contrast, an initial presence of the band and a progressive growth would indicate that the initial ice sample is made of a mixture of crystalline and amorphous ice and that the amorphous ice crystallises as temperature increases.

During this experiment, the spectra of water ice were recorded from 1.302 to 1.800 µm in the temperature range 96 to 138 K. Each spectrum took 6 minutes to acquire. From the beginning, the absorption feature of crystalline water ice is present (Figure 12, B) at its maximum depth observed during the entire sequence. The first three hours of the experiment show a strong decrease of the band strength followed by two regimes: first, a very fast decrease of band strength (first 7 cubes, ~45 minutes) followed by a slower decrease. In between, a sudden small increase of band strength (by ~5%) is noted as the temperature of the sample reaches ~117K. After three hours, both the 1.65 µm band strength and the sample temperature stabilise at values of 0.5-0.55 and ~140 K, respectively.

Our interpretation of these observations is that the initial pure SPIPA-B granular ice sample is mainly crystalline, formed of $I_h$ ice. A low amount of amorphous ice (shell and/or contamination) is also present initially and its crystallisation results in the sudden increase of the 1,65 µm band depth observed as the temperature reaches ~117K (Figure 12, B). The evolution of the band depth then follows the evolution of the temperature of the sample.

The situation is probably different with salt solutions, as the presence of the salts inhibits the ice crystallisation. As for pure ice particles, salty droplets solidify from the outside to the inside, making the ions migrate within the particles. This process results in inner veins and cracks within spherical particles to be filled with a vitreous solidified water, enriched in dissolved ions, and surrounded by a mantle of pure crystalline ice. As for pure ice particles, it is also possible that a shallow shell of amorphous ice also exists at the surface of the particles, produced at the first contact of the liquid droplet with the liquid nitrogen.

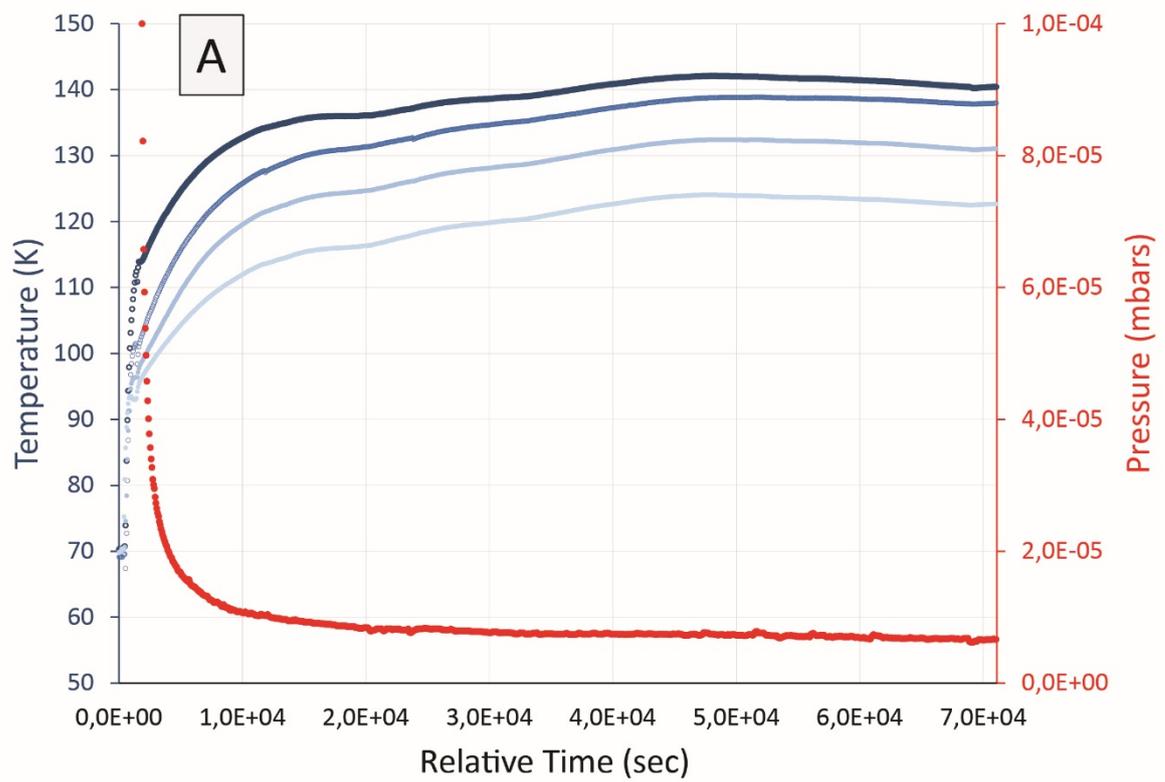
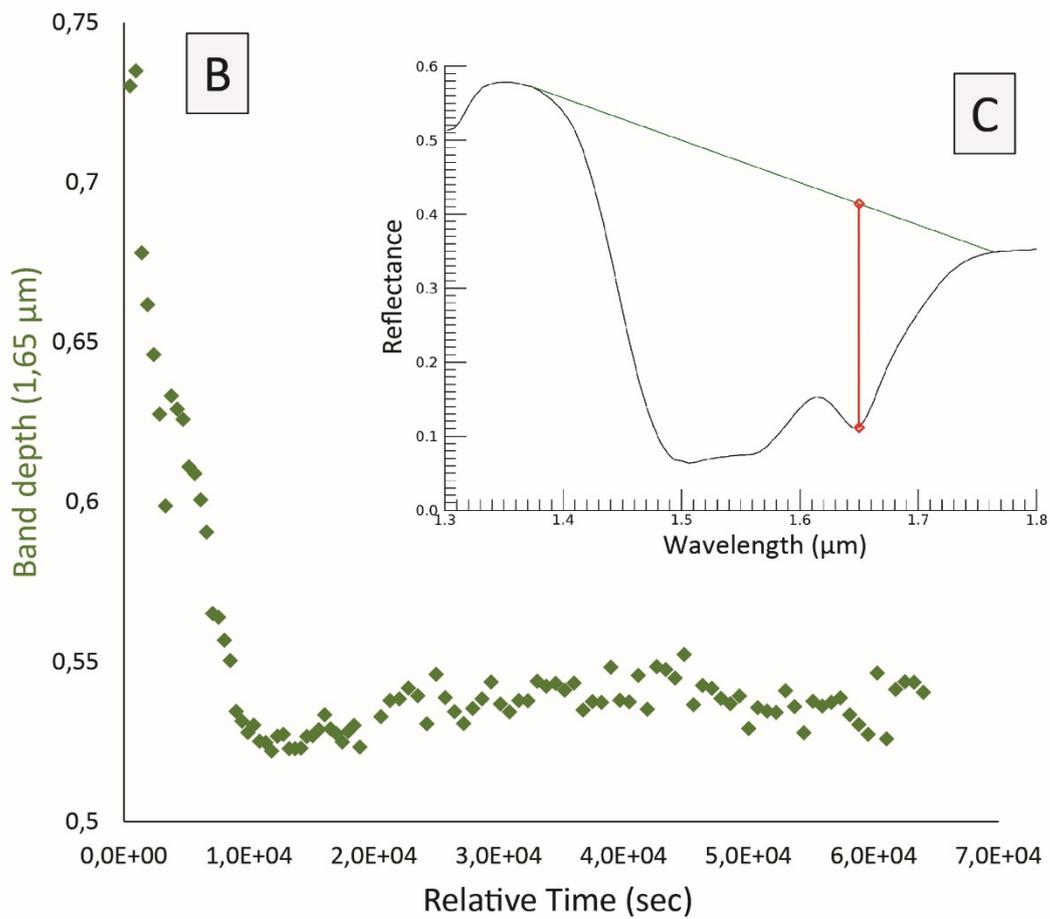

Figure 12: Results of an experiment performed to assess the crystallinity of the water ice particles produced with SPIPA-B. A) The red curve represents the pressure evolution within the SCITEAS-2 chamber during the 17 hours of *the experiment. The four blue curves represent the temperatures recorded by four sensors placed at different depths within the ice sample. The darkest curve is from a sensor positioned slightly above the surface of the sample at the beginning of the experiment. The lighter curve is measured by the deepest sensor within the sample. The second darkest curve is most likely the more representative sensor by being within the ice and close to the surface (less than a centimetre). B) Evolution of the depth of the 1,65-µm band through time (green). The sudden increase of 5% in the band depth value occurs between the 7$^{th}$ and 8$^{th}$ cube. C) Definition of the depth (red) of the 1,65 µm band. The wavelengths in the NIR were chosen over the entire absorption complex of the water ice around 1.5 µm (1,374 to 1,764 µm).*

## 4.2. Properties of ices from brines, comparison between slabs and granular ices

Following the analysis of spectral changes between the initial and final state of each sample, we now study the evolution of selected spectral criteria quantitatively and continuously throughout the sublimation sequence. Figures 13, 14 and 15 show the evolution of the Normalised Integrated Band Area (NIBA) and its comparison to the barycentre of the bands (referred to as absorption band centroid) at 1.5 and 2.0 µm.

Changes to the water ice absorption bands have been documented for salty hydrates. They have been studied by Raman spectroscopy, repeatedly showing changes in wavenumber and shape of the absorption compared to pure water ice (Hamilton and Menzies, 2010; Shi et al., 2019; Thomas et al., 2017; Wang et al., 2006; Yeşilbaş et al., 2018).

A slow crystallisation following the thermodynamic equilibrium produces the salt species with the highest degree of hydration. The sublimation of these samples leads to spectra which always remain dominated by the spectral features of water, even after sublimation for long periods of time (See Figure 5, 6 and 7 for $MgCl_2$, $MgSO_4$ and $Na_2SO_4$ slabs, respectively). This is reinforced by the fact that the slabs offer longer optical paths for the water ice, maximising the strength of its absorptions at the beginning of the sublimation sequence. After the sublimation, the absorption bands of the water-of-hydration in the hydrated salt dominate.

Compared to pure water, the spectral features of water in hydrated salts (as stoichiometric compound) are distorted because of the interaction between the water molecules and the ions in the crystal. The stretching (1.5 µm), bending (2.0 µm) and combination of both (1.25 µm) modes of the $H_2O$ molecule are slightly modified. The length of the hydrogen bond is dependent on the arrangement of the crystal lattice as explained in (Shi et al., 2019) for MgCl – $H_2O$. For each hydrate ($MgCl_2 \cdot nH_2O$, with n = 2, 4, 6, 8, 12), they observe different lengths of the O-H bonds within the basic unit of the crystal structure ($Mg^{2+}$ and its atomic neighbourhood within its octahedral organisation). Starting with n = 6, the octahedrons of $[Mg(H_2O)_6]^{2+}$ are no longer linked to each other. With fewer water molecules, the $Cl^-$ connect to each other to construct chains. For n = 8 and 12, in addition to the water molecules linked to the metal ions, *"free (non-octahedra-coordinated)"* water molecules result in a variety of O-H bond lengths (n = 8 or 12 having respectively 8 or 16 free water molecules).

(Wang et al., 2006) studied the hydration states of $MgSO_4$ with Raman spectroscopy and also concluded that the O-H bond length varies with the hydration state ($MgSO_4 \cdot nH_2O$, with n = 1, 2, 3, 4, 5, 6, 7). A similar study was later performed for sodium sulphate hydrates ($Na_2SO_4 \cdot nH_2O$, with n = 7, 10) by (Hamilton and Menzies, 2010) showing Raman spectra affected by shifts of vibrational frequencies compared to pure $H_2O$, indicating variability of O-H bonds lengths within the crystal. For hydrated NaCl, the studies from (Baumgartner and Bakker, 2010; Samson and Walker, 2000; Thomas et al., 2017; Yeşilbaş et al., 2018) show Raman spectra of *hydrohalite* with variations of the peaks attributed to the presence of water-of-hydration by comparison with pure water crystals. This leads to the same interpretation that the presence of the salt alters the O-H bonds and modifies the vibrational frequencies compared to pure water crystals.

The absorption bands related to the water molecule that are seen in the spectra of analogues are therefore displaced and distorted compared to pure water ice due to the level of hydration as well as the nature of the salt.

### 4.2.1. Evolution of the reflectance of salty slabs during the sublimation

Both the phase diagram of each compound and the cooling kineticsdetermine the number of hydrates formed. The more time spent on the solidus of a hydrate during the cooling of the initial solution, the more hydrates are formed because both temperature and concentration are at the right conditions for a longer time. The final phases formed are well crystallised and separated. The initial spectra of the salty slabs are influenced by the first few hundred micrometres below the surface and correspond to a superimposition of water ice and hydrates. When highly hydrated salts such as for $MgSO_4$, $Na_2SO_4$ and $MgCl_2$ form the spectra are strongly dominated by water, the absorption complexes at 1.5 and 2.0 µm are flattened and broadened, and all other spectral features are barely recognisable. For NaCl, which forms only *hydrohalite* (2 water molecules) as stable hydrate, the spectra are dominated by water ice, but the absorption features specific to *hydrohalite* are identifiable from the beginning and at every concentration (Figure 4).

The sublimation progressively removes water from the top layers of the slab, firstly pure water ice crystals and then water from the salt hydrates, leading to partial dehydration of the hydrates. A good example of such an evolution is the NaCl experiment (Figure 4). The final spectrum exhibits localised absorption features related to the *hydrohalite* and seems depleted of pure water ice crystals. Extending this conclusion to the others salty slabs ($MgSO_4$, $Na_2SO_4$ and $MgCl_2$), after a few hours of sublimation, all surfaces are dominated by the spectral signatures of the hydrates. The water of hydration is the dominant compound due to the highly hydrated stable crystalline forms. The conditions of low pressure and temperature affect this water of hydration, leading to the dehydration of the crystals. The removal of this water, strongly bonded to the metal ions within the crystal lattice, is a much slower process. Figure 13 regroups the evolution of the NIBA and the centroids of the bands during the sublimation experiments for slabs of $MgCl_2$, $MgSO_4$ and $Na_2SO_4$. The NaCl slab shows specific absorption features different from the 1.5- and 2.0-µm absorption complexes. Its evolution is therefore shown separately in Figure 14.

The wavelengths range used for the calculation varies between different chemistries ($MgSO_4$, $Na_2SO_4$ and $MgCl_2$, NaCl), but remains the same for a given composition at different concentrations (e.g. for NaCl at 10, 20 and 30 wt% for example). This selection has been tailored to the spectral behaviour of each analogue due to the strongly flattened and broadened absorptions within slabs. The 1.5- and 2.0-µm absorption complexes have been chosen as they are the most affected by the presence of hydrates and in some cases, the only absorption bands exploitable. The shallower spectral features described in Section 3.1 provide information about the nature of the hydrates but are not sufficiently pronounced to perform accurate computations of the NIBA. By computing the NIBA over a wide range of wavelengths, we risk inclusion of extraneous bands. Nevertheless, the wavelengths have been chosen to mitigate the overlap of absorption features. The increase or decrease of the NIBA, as well as the shifts of the centroid positions provide first-order quantitative insights into the evolutions of the physical properties of analogues.

### 4.2.1.1. NIBA vs Centroid of $MgCl_2$ slabs

The $MgCl_2$ slabs display stable 1.5-µm complexes (Figure 13, A). Stable barycentres through the experiment indicate that the sublimation removes water ice without dehydrating the salts at the surface of the sample. As seen in Figure 13 a), the 25 and 33 wt% $MgCl_2$ slabs display slight oscillations of the band centroid (4 to 6 nm) but no clear displacement of the band. The NIBA values are remarkably stable. The number of points and their distribution give an idea of the relative error generated by the computation, which seems to be within a few per cent. The 12.5 wt% sample does not show any displacement of the band or change of its intensity.

The 2.0-µm complex shows, for all three concentrations (Figure 13, B), an increase of the intensity of the band. It shows a transition from a band almost saturated by the presence of water ice to a band shaped by the hydrates. The centroid shifts slightly toward longer wavelengths, within 10 nm for the 12,5 and 33 wt% cases, and roughly 40 nm for the 25 wt% case. These absolute values of centroid positions have to be taken with precaution because of the saturation of the water band, complicating the quantification of the band's evolution through sublimation.

The sublimation of water ice leads to the appearance of absorption bands related to the hydration of the salt, but the dehydration of hydrates of $MgCl_2$ has not been detected.

### 4.2.1.2. NIBA vs Centroid of $MgSO_4$ slabs

The 10 and 20 wt% slabs of $MgSO_4$ (Figure 13, C) show similar behaviour at 1.5 µm: first, a decrease of the NIBA (around 50%: from 0,308 to 0,145 for the 10 wt% and from 0,305 to 0,165 for the 20 wt%) followed by a shift in the centroid position by 8 and 6 nm, respectively. The 10 wt% case displays a clear transition between the decrease of the NIBA and the shift of the centroid, whereas the 20 wt% case shows evolutions of the two parameters simultaneously. The 30 wt% case also shows a diminution of the intensity of the band as well as a short displacement to longer wavelengths (8 nm). These observations are linked to the sharpening of the band due to the sublimation of water ice. In the 30 wt% case, the growth of the 1.62-µm and, to a lower extent, of the 1.77-µm absorption feature displaces the centroid of the bands toward longer wavelengths. This 30wt% saturated case exhibits a different behaviour, the final spectra being similar to the ones obtained by (Dalton et al., 2005; McCord et al., 2001), for *starkeyite*, *pentahydrite* and *hexahydrite* (respectively $MgSO_4 \cdot [4, 5, 6]H_2O$) or $MgSO_4 \cdot 12H_2O$ around 100K (Dalton et al., 2005). In the latter article, the effect of the grain size on the absorption feature at 1.69 µm is also showed, which increases the intensity of this band ($MgSO_4 \cdot 5H_2O$ in their fig. 2). Here, the growth of this feature shows the role of the salt concentration. Following the phase diagram of $MgSO_4$, saturation happens with 27 wt% at ambient conditions (1 bar, 293 K). With 30 wt% concentration, crystals of *epsomite* ($MgSO_4 \cdot 7H_2O$) are formed in the solution. The evolution of the spectral criteria shows that dehydration affects every concentration and every analogue with the same efficiency. Therefore, an explanation could be the nature of hydrates present before the sublimation. The water is firstly removed by sublimation of pure ice, uncovering the hydrates present within the sample. By comparison with Fig. 1B from (Dalton and Pitman, 2012), our 30 wt% exhibits a 1.69-μm absorption feature similar to their sample of epsomite with a grain size between 100 and 200 micrometres at 120 K.

The absorption band at 2.0 μm (Figure 13, D) shifts toward shorter wavelengths for all concentrations. The higher is the concentration, the shorter the shift is. The 10 and 20 wt% cases are both marked by a diminution of the intensity of the band. With the exception of the first four hyperspectral cubes, all other points cluster around NIBA ~0.23. The interpretation of such behaviour is as previously described: the sublimation removed the ice to exhibit the spectral signatures of the hydrates. The 10 wt% and 20 wt% cases display sharpened absorption bands. This latter concentration also exhibits a shallow feature at 2.06 μm, which reduces the shift of the centroid to shorter wavelengths. For the 30 wt% case, as the 2.0-μm band is flattened and the 2.06-μm feature grows during the sublimation, the shift is even smaller. The absorption band at 2.0 μm starts to lose intensity because of the loss of water ice during the 3 to 4 first cubes (2 hours of sublimation), to finally increase because of the growth of the 2.06-μm feature. Its position coincides with an absorption feature of *epsomite* (Dalton and Pitman, 2012).

### 4.2.1.3. NIBA vs Centroid of $Na_2SO_4$ slabs

The 30 wt% is a saturated case (McCarthy et al., 2007). The 1.5-μm band is weak, and the 2.0-μm band never appears (Figure 7). From the NIBA computation (Figure 13, E and F), the 15 wt% case, close to saturation, shows a cluster of points, without a noticeable shift of the centroid position. For this concentration, the dehydration process that would lead to the appearance of specific absorption bands is not noticed.

The 3 wt% case is the only one showing a distinctive behaviour. In the case of the 1.5 μm band, the sublimation results in a decrease of the 1.5 μm band for the first 4 to 5 cubes (about 2 hours of sublimation) and then the centroid shifts toward shorter wavelengths (as show by the green arrow, Figure 13, E). In the case of the 2.0 μm band, the intensity increases, as the almost saturated band transforms into a well-defined band, and shifts toward shorter wavelengths. Its position and shape, with an almost flat bottom between 1.94 and 2.0 µm, are diagnostic of the *mirabilite*, as seen by (Dalton, 2007; Dalton et al., 2005).

Evolution of the NIBA vs. the band centroid for the slabs (MgCl$_2$, MgSO$_4$, Na$_2$SO$_4$)

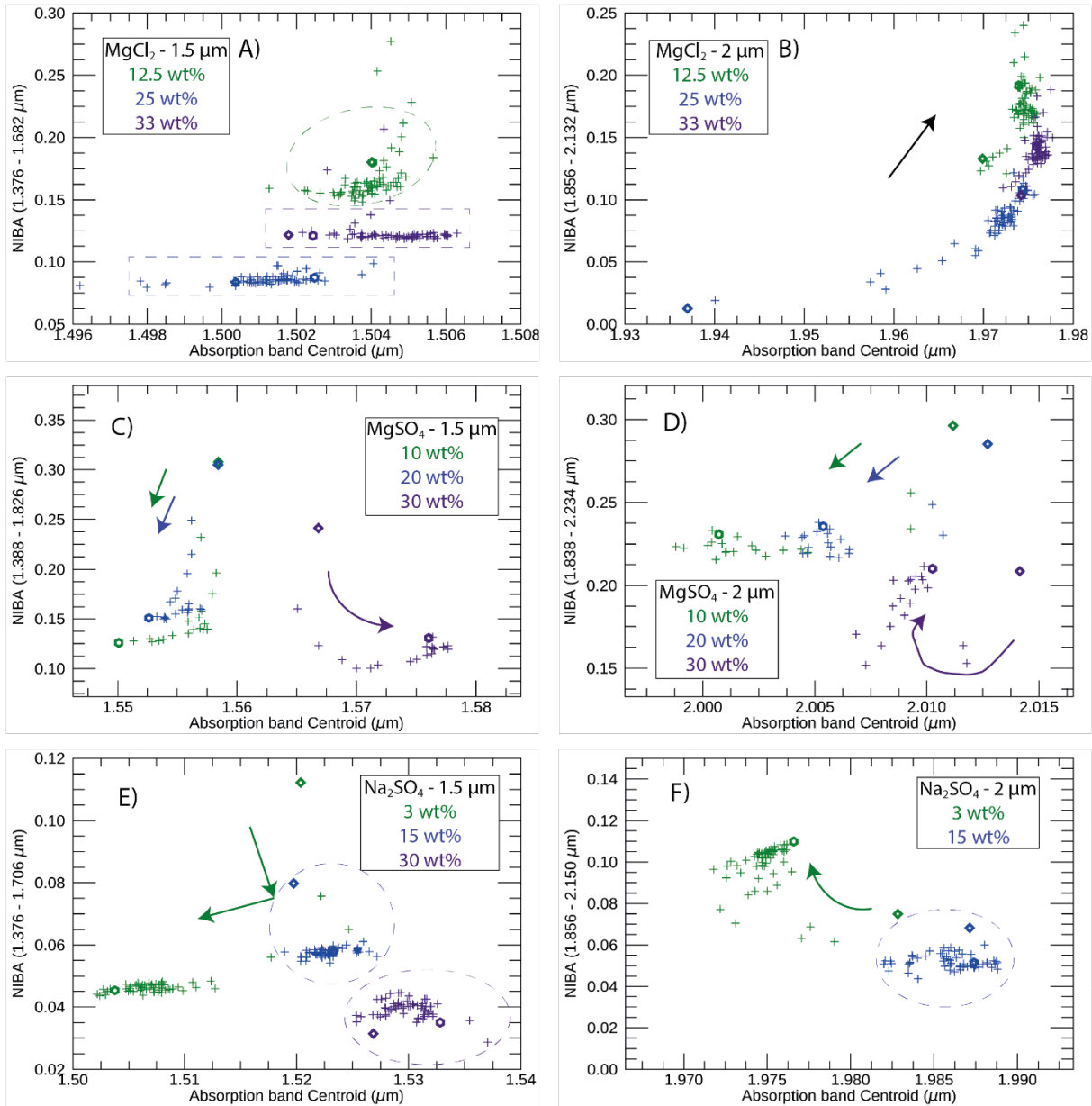

*Figure 13: Evolution of the NIBA vs. the band centroid during the sublimation experiment for slabs at 1.5 and 2.0 µm. The wavelengths used for the computation of the NIBA are indicated as the title of the y-axis. The NIBA values of water slabs have not been plotted because the bands at 1.5 and 2.0 µm are completely saturated. The coloured arrows are indicating the evolution of the NIBA and centroid values for specific concentrations. The black arrow on B indicates the trend for all the concentrations. The diamonds and hexagon symbols are respectively indicating the initial and final hyperspectral cube. A, C, E show the evolution of the 1.5-µm complex. B, D and F show the evolution for the 2.0-µm complex. Dashed circles indicate cases where the values do not show clear evolution but rather clusters of points.*

### 4.2.1.4. NIBA vs Centroid of NaCl slabs

The evolution of the NaCl slabs through sublimation differs from the ones of the other salts because of the unicity of level of hydration of the NaCl hydrate. As mentioned previously, the *hydrohalite* (NaCl·2H$_2$O) is the only stable hydrate formed, which is a low hydration state compared to the other salts used in this study. Therefore, all concentrations tested led to a similar final spectrum, without saturated absorption bands.

The absorption band at 1.5 μm evolves towards two different double absorption bands in 3 to 4 hyperspectral cubes (around 2 hours of sublimation). Their evolution is presented in Figure 14 a) and b). As water ice sublimates, the absorption band at 1.5 μm weakens, revealing the *hydrohalite* signature at 1.8 μm, which gradually intensifies. Both bands shift toward shorter wavelengths: the displacement of the complex at 1.5 μm (by 60 nm for the 10 and 20 wt%, and by 30 nm for the 30 wt%) is larger than the one at 1.8 μm (4 to 6 nm). The 1.5 μm band emerges from the water ice band (Figure 14, a)) instead of being exclusively related to the hydrate, as the 1.8 μm band.

The absorption complex at 2.0 μm also splits into two bands (Figure 14, C and D): a double one at 1.95 μm and a simple one at 2.17 μm, related to the hydrates. As the amount of water ice decreases, the intensity of the hydration signature increases. The displacement of the band (by 30 to 40 nm) indicates the growth of a band at a wavelength where, before sublimation, only a small feature was identifiable. The double band at 1.95 μm shifts slightly (by ~5 nm) toward longer wavelengths and its intensity decreases as the band sharpens due to the removal of water ice.

The sublimation removes water ice, so that the bands of the hydrates appear more intense, but the *hydrohalite* is not affected by dehydration; which would make the absorption bands disappear. As depicted in Figure 14, the first cubes are marked by a strong evolution of the NIBA due to a strong evolution of the values of reflectance because of sublimation. The rest of the experiment is characterised by bands at roughly the same position and intensity. The total removal of water would have led to a completely flat spectrum, as the NaCl has no absorption bands within the VIS or NIR region. Hence this confirms the presence and sustainability of *hydrohalite* through sublimation.

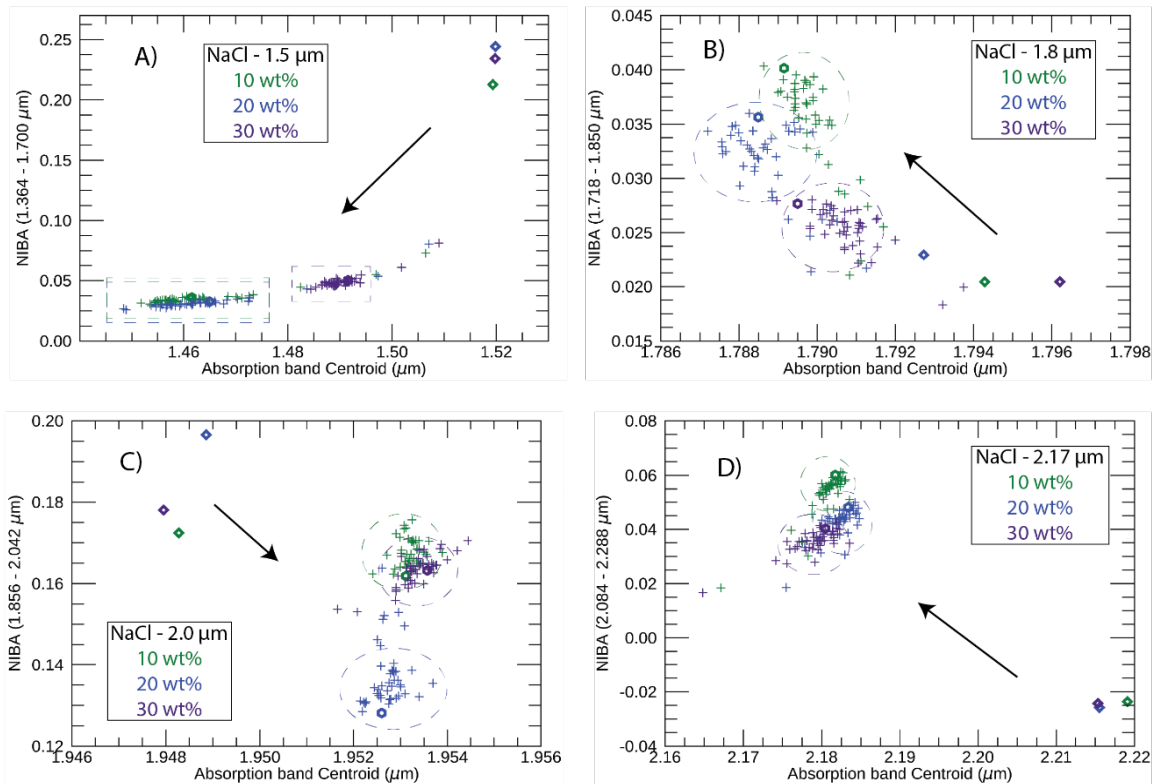

*Figure 14: Evolution of the NIBA for NaCl slabs from the initial 1.5- and 2.0-μm absorption complexes vs the band's centroid position. The wavelengths used for the computation of the NIBA are indicated as the title of the y-axis. The NIBA values of water slabs have not been plotted because the bands at 1.5 and 2.0 μm are completely saturated. A and B show the evolution of hydrohalite bands within the 1.5-μm complex. C and D show the evolution of bands within the 2.0-μm complex. The black arrows indicate the trend of the evolution of the NIBA and centroid position with sublimation. The diamonds and hexagon symbols are respectively indicating the initial and final hyperspectral cube.*

From the examples discussed in this section, the sublimation of water ice (not hydration water) for salty slabs analogues is fast. In each case (Figure 13 and 14), the samples exhibit a noticeable change of the reflectance values within 1 to 2.5 hours. The evolution of the bands is related to the hydrates present and occur within the first 2 to 3 hours of desiccation. The spectra remain nevertheless dominated by absorption bands of water because of its presence within water-rich hydrates ($MgCl_2 \cdot 12H_2O$; $MgSO_4 \cdot 11H_2O$; $Na_2SO_4 \cdot 10H_2O$, Figure 5, 6, 7, and Figure 9, 10 and 11). These observations show that sublimation and dehydration would have to continue for longer periods of time to remove completely the water from the analogues under the pressure and temperature conditions used for this study. Even reducing considerably the amount of water would not result in a complete vanishing of the absorption features of water.

Despite their low hydration states, $MgCl_2$ and $MgSO_4$ still display strong signatures of water (Crowley, 1991; Dalton, 2007; Dalton et al., 2005; Hanley et al., 2014; McCord et al., 2001; Shi et al., 2019). Finally, having the hydrates mixed with water ice also differs from having the hydrates alone, therefore complicating the identification and the accuracy of spectral criteria computation. One can note that with a low amount of hydration, such as for the NaCl slabs, the absorption bands are significantly less intense, because of the lower amount of water, which facilitates the identification of well-resolved hydrate spectral features.

### 4.2.2. Evolution of the reflectance of salty granular ices submitted to sublimation

In the case of the flash-frozen salty ice particles (Figure 8, 9, 10 and 11), the amount of hydrates created is lower than in the case of compact slabs due to the higher cooling rate. The general behaviour observed with all granular ice samples is that the higher the salt concentration in the solution is, the stronger the absorption features are. Still, the amount of hydrates always remains low compared with slabs and the disorder induced by the flash freezing leads to the formation of a glass-like phase in the interior of the grains (see the discussion in Section 4.1).

An overall trend can be derived from the observations of every salt and every concentration: the intensities of the absorption complexes at 1.5 and 2.0 µm decrease during the experiments as a result of ice sublimation (Figure 15). With the exception of the 1.5-µm absorption band of $Na_2SO_4$, all the bands shift to shorter wavelengths.

In Figure 15, data points for pure water ice samples are also shown (red dots). Similar values were shown for the compact slabs in Figure 13 and Figure 14 as pure ice slabs are too transparent, and the signal is polluted by the presence of the sample holder visible through the ice. Note that in Figure 15 the points corresponding to pure water ice move between the different plots. This is because each plot uses slightly different wavelengths to calculate the NIBA and centroid values. Using the same spectral criterion, all centroids for pure water ice are within +/- 0.005 nm and NIBA values within 0.1.

### 4.2.2.1. NIBA vs Centroid of NaCl granular ices

For the NaCl granular ices (Figure 15, A and B), the absorption complexes at 1.5 and 2.0 µm disappear progressively. Compared to the slab, the spectral signatures of the hydrohalite are reduced due to the lower amount of hydrates formed, leading to reduced values of NIBA as well as a noisy distribution of the band's centroids. The spectrum of the surface evolves from an ice spectrum toward a pure NaCl spectrum (Figure 3) with adsorbed water.

### 4.2.2.2. NIBA vs Centroid of MgCl$_2$ granular ices

The granular ices produced from solutions of MgCl$_2$ display spectra (Figure 15, C and D) marked by several absorption bands diagnostic of the hydrates of MgCl$_2$. A common behaviour to both the 1.5- and 2.0-µm band is that the higher the salt concentration in the initial solution is, the longer the shift in wavelength is (20 and 25 nm for the 12.5 and 33 wt% cases, respectively). In granular ices produced through flash freezing, the higher the amount of salt, the stronger the absorption features related to the hydrates. Even with a fast cooling rate, one can notice that a small but significant fraction of hydrates forms.

The hydration level of the hydrates depends on the concentration and is depicted in the phase diagram of MgCl$_2$ – H$_2$O (Davis et al., 2009; Li et al., 2016). With concentrations lower than 21 wt%, the cooling induces the formation of water ice, then only under 240 K, MgCl$_2$·12H$_2$O starts to form. Between 21 and 31.5 wt%, the formation of MgCl$_2$·12H$_2$O occurs already at 256 K, and it remains the only hydrate producible. For concentrations higher than 31.5 wt%, $MgCl_2·8H_2O$ can form and if the concentration reaches 34 wt% or higher, $MgCl_2·6H_2O$ can form.

The exact internal location of each phase is difficult to determine; a recent study by (Fox-Powell and Cousins, 2021) suggests a complex distribution within cracks and veins within the particles. As water tends to expel salt from its lattice as it crystallises, the salt concentration of the remaining liquid progressively increases. Part of this highly concentrated liquid phase can therefore form some hydrates even when the salt concentration in the initial solution is not normally sufficient for their formation. This is a form of fractional crystallisation (see (Zolotov and Shock, 2001)) and is a probable explanation for the presence of spectral features related to lower hydration states with a higher concentration in $MgCl_2$ – $H_2O$ granular ices.

The absorption feature located around 1.75 µm (Figure 9) is more pronounced as the hydration level is low (Li et al., 2016), being related to $MgCl_2·nH_2O$ with n = 6 and lower. A possible explanation for the presence of this feature in the case of the 33 wt% sample (and to less extent with the 25 wt% case) would be the increase of salt concentration in the last liquid phase within the interior of particles. It would suggest an increase of the concentration of the order of 1 to 9 wt% (for 33 and 25 wt%, respectively) to reach the 34 wt% required for the formation of $MgCl_2·6H_2O$.

The distortion and displacement of the water features are stronger with samples that are more concentrated in salt. The sharpening of the band at 1.5 µm is stronger with higher concentrations (or lower hydration state for the present case) leading to a more symmetric band and a centroid positioned at lower wavelengths.

The absorption complex at 2.0 µm appears similar for every concentration. It shows an absorption feature at 2.0 µm related to water ice, within the absorption band at 1.96 µm (1.97 µm for (Shi et al., 2019)). The transition from a spectrum dominated by water ice to a spectrum dominated by hydrates is responsible for the observed shifts in wavelengths. Moreover, one can notice that the points from the last hyperspectral cubes cluster, meaning that after removing the ice, the hydration state is not strongly modified.

### 4.2.2.3. NIBA vs Centroid of MgSO$_4$ granular ices

The MgSO$_4$ – H$_2$O granular ices show 1.5- and 2.0-μm features (Figure 15, E and F) which also shift toward shorter wavelengths with sublimation. The sublimation of water ice is more noticeable for lower salt concentrations, resulting in a stronger decrease of the NIBA (around 0.43 to 0.2, 0.12 and 0.08 at 1.5 μm and around 0.525 to 0.32, 0.24 and 0.19 for the 30, 20 and 10 wt%, respectively) due to the higher proportion of water.

The shift in wavelength is stronger for lower concentrations. The 1.5-μm band flattens more for lower concentrations, as the hydration level is lower with higher concentrations. The 10 wt% MgSO$_4$ is expected to form only the *meridianiite* whereas the 30 wt% case is saturated and is expected to form a fraction of *epsomite*. As described in Section 3.1, the feature around 1.76 μm is indicative of *epsomite*. Its presence in the final hyperspectral cube of the 20 wt% sample suggests an increase of salt concentration in the interior of the particles resulting from the flash freezing procedure. *Epsomite* within the analogues leads to a marked "V" shape of the 1.5 μm complex instead of the "square root" shape expressed in the 10 wt% case.

The 2.0-μm complex evolves towards "V" shaped bands, with a flatter floor of the bands at higher salt concentrations as shown by (Wang et al., 2011), reducing the shift to shorter wavelengths of the centroid with higher concentrations.

The comparison between the MgCl$_2$ and MgSO$_4$ – H$_2$O granular ices reflects two different behaviours regarding the salt concentration that are well expressed with the spectral criteria chosen. The strong absorption features produced by lower hydration states within MgCl$_2$ – H$_2$O granular ices generate evolutions of the bands comparable to the evolution of higher hydration states of MgSO$_4$ – H$_2$O granular ices.

### 4.2.2.4. NIBA vs Centroid of $Na_2SO_4$ granular ices

The granular ice with 3 wt% of $Na_2SO_4$ only shows an overall diminution of the NIBA values without shift of band positions (Figure 15, G and H). This analogue mimics the spectral features of pure water ice, with increased reflectance values due to $Na_2SO_4$ salt powder at the surface.

The two samples with 15 and 30 wt% concentrations display the absorption features of *mirabilite* which sharpen the 2.0 µm complex while sublimation proceeds and progressively shifts the bands toward slightly shorter wavelengths. The absorption feature at 1.77 µm does not affect the 1.38- – 1.70-µm range. The flat and broadened absorption complex at 1.5 µm is due to the presence of $Na_2SO_4 \cdot 10H_2O$. During the sublimation, the band loses intensity but remains flatter than in the 3 wt% concentration case, therefore inducing only a small shift in wavelength.

In the compact slabs, our experiments did not result in significant dehydration of the salt hydrates, and the main process at play is the sublimation of the water ice. Granular ices are even more sensitive to sublimation because their surface area to volume ratio is higher, liberating excess water (not linked to the hydrate crystal structure) more efficiently. Moreover, the amount of water not bound as a crystal lattice component is higher in granular particles, which in turn increases the amount of excess water to sublimate.

An important thing to notice when systematically comparing the pure water case produced for each experiment (in red in Figure 15) to the salty analogues is that with sublimation, the pure water case always maintains the same values of NIBA and the centroid position. The presence of the salt, even in small quantity changes the absolute values of reflectance even if it does not affect the spectral behaviour of the sample significantly (see, $Na_2SO_4$ – $H_2O$ granular ices at 3 wt%, Figure 15 G and H).

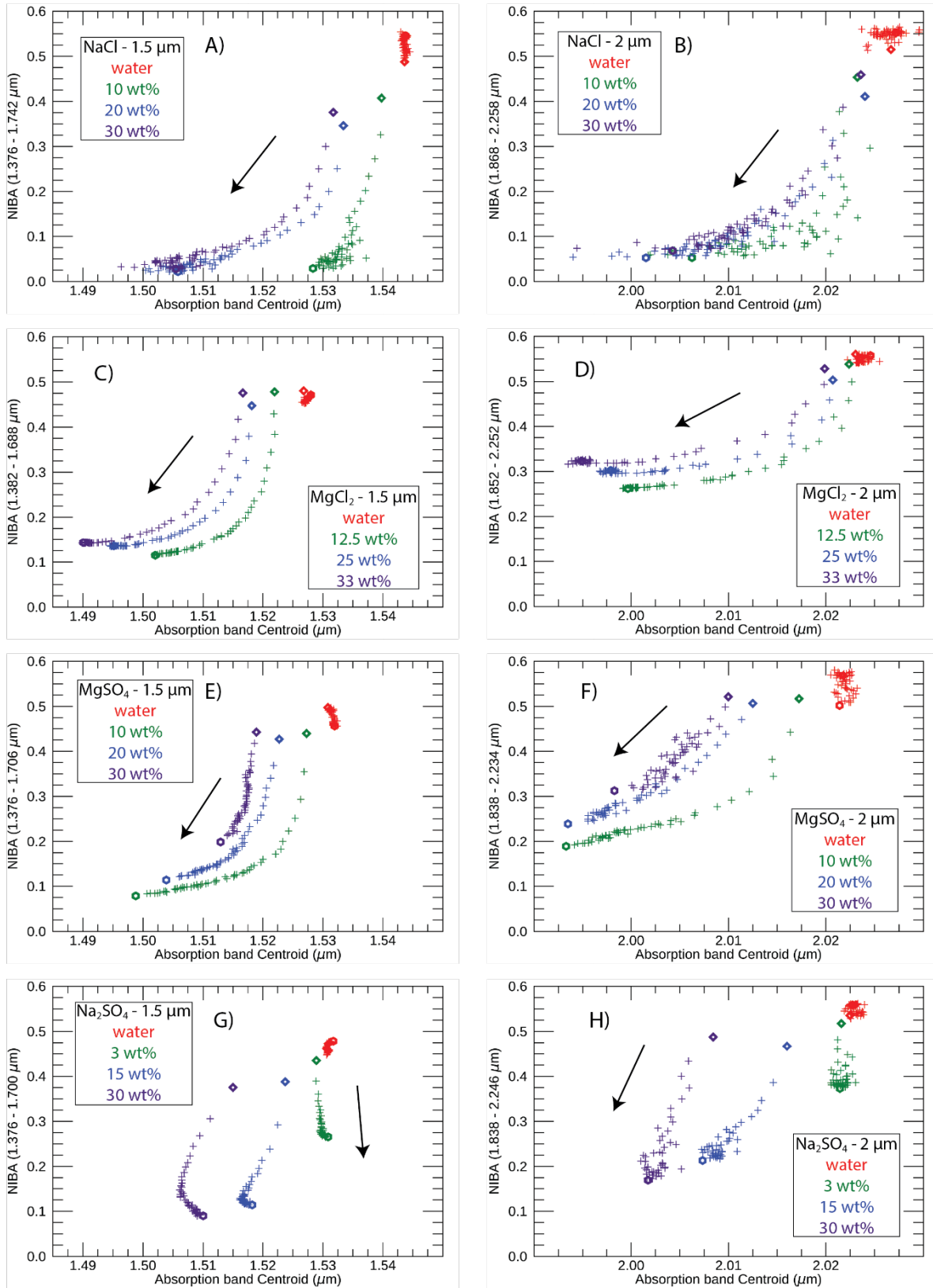

*Figure 15: Evolution of the spectral criteria NIBA for the granular ices (SPIPA-B) at 1.5 (left plots) and 2.0 µm (right plots) vs. the centroid position. Each time, for the different chemistries, the criterion has to be adapted in order to record the evolution of the absorption complex related to water. The wavelengths taken for the computation are indicated as the title of the y-axis. This also affects the pure water ice bands which are therefore slightly*

*shifted in position between the plots. The diamonds and hexagon symbols indicate the initial and final hyperspectral cube, respectively. The black arrow describes the general trend of the evolution of the NIBA values with time. A, C, E and G show the evolution of the 1.5 µm absorption complex. B, D, F and H show the evolution of the 2.0 µm complex.*

### 4.2.3. Slab versus Granular ices

The decrease of the NIBA is, in most cases, 5 to 10 times higher with granular ices than with compact slabs. The amount of hydrates generated with slabs is higher as well as their degree of hydration. With the flash freezing procedure, the amorphous part of the particles is made of a glassy mixture of salt and water but weakly bonded to each other compared to the slabs. The removal of these water molecules through sublimation is easier than the removal of water-of-hydration and is enhanced for granular particles due to their higher surface-to-volume ratio.

The trends observed with the spectral features of hydrates are explainable by their production procedure. Dehydration of the salts may occur, but not as a dominant process compared to the sublimation of water ice. The hydrates form during the sample production procedure, and as the sublimation removes the water ice, it makes the hydrates visible and identifiable.

The removal of pure water ice, as well as salt dehydration when it happens, leads to modifications of the structure of the icy analogues, inducing cracks, increasing the number of small scatterers, which increases the values of reflectance due to internal scattering. This behaviour has been seen with every sample produced for this study. It has to be noted that the average size distribution of scatterers within a slab is much larger than for granular ices. With a larger grain size, the pathway of photons inside the grains increases, and thus increases the absorption, which explains the deeper spectral features observed with slabs.

## 4.3. Comparison between intimate mixture analogues produced with current database of experimental work on salty hydrates

We now compare the reflectance data of pure salt hydrates available in the literature and the analogues produced in this study. The aim is to assess the similarities and differences regarding the different parameters available, such as the concentration, the grain size and the production method. We begin by summarising the findings from previous studies.

### 4.3.1. Previous studies on sulphate bearing species ($SO_4^{2-}$)

(Orlando et al., 2005) studied the reflectance of mixed salty brines of $MgSO_4$, $Na_2SO_4$ and $H_2SO_4$ in various proportions in order to identify spectrally the non-icy components in Galileo Near Infrared Mapping Spectrometer data (NIMS, (Carlson et al., 1992)). The flash-freezing procedure coupled with thermal cycles to dehydrate the analogues produced "V" shaped 2.0-µm complexes as well as a 1.6 – 1.8 µm region marked by several weak absorption features. As sulphuric acid is present within their analogues, some absorption features they recognize can be attributed to this chemical compound. We have not used $H_2SO_4$ in this study. Nevertheless, the "V" shaped 2.0-µm complex is in good agreement with our flash-frozen analogues produced with $Na_2SO_4$ and $MgSO_4$, especially after sublimation, as well as our slabs with 10 wt% $MgSO_4$ (Figure 6). Their spectra do not exhibit the small band at 1.77 µm, seen in our 15 and 30 wt% $Na_2SO_4$ – $H_2O$ granular ices (Figure 11), nor the feature within the 1.5-µm absorption complex for $MgSO_4$ – $H_2O$ granular ices.

(McCord et al., 2002) produced flash-frozen ice particles from brines of saturated solutions of $MgSO_4$ and $Na_2CO_3$ with particles sizes estimated at 250 µm, and $Na_2SO_4$ icy particles with ~1mm grain size. Their first spectra of $MgSO_4$ and $Na_2SO_4$ display strong absorptions at 1.5 and 2.0 µm, in good agreement with our slab experiments. The thermal cycling applied to their samples results in a removal of water by dehydration. Our experiments did not lead to such dehydration. Their final spectra show sharp absorptions at 1.5 and 2.0 µm, which is different from our observations of sublimated slabs. These spectra look much more similar to our flash-frozen granular ices that show stronger effects of water loss.

(Dalton, 2003) worked on magnesium sulphate hydrates, producing NIR spectra of $MgSO_4 \cdot nH2O$ (n = 1 to 7). The *meridianiite* and *mirabilite* spectra are depicted in (Dalton et al., 2005) from 50 to 250 K in steps of 50 K. At temperatures below 100K, their spectra display sharper absorption features that were not observed within our analogues.

### 4.3.2. Previous studies on chlorides-bearing species (Cl-)

(Hanley et al., 2014) produced hydrated salts of $MgCl_2 \cdot nH_2O$ (n = 2, 4 and 6) and measured their reflectance spectra at 298 and 80 K. As the temperature decreases, the spectral features are better resolved. The features present in the 1.5- and 2.0-µm absorption complexes of their hydrates at 80K have never been observed within the icy analogues produced for this study. This is linked to both the temperature - which was several tens of K higher in our experiments - as well as the mixture of salt and water, which prevents the observations of such well-defined absorption features. Except for these features, our $MgCl_2$ – $H_2O$ granular ices after sublimation (Figure 9) display absorptions similar to those shown by Hanley et al., (2014).

(Shi et al., 2019) performed a dedicated study of $MgCl_2 \cdot nH_2O$ compounds, providing an exhaustive description of the combinations of overtones of water molecule vibrations responsible for the absorption bands (asymmetric and symmetric stretching, bending as well as metal – OH mode). Their VNIR measurements are in good agreement with the ones from (Hanley et al., 2014) at 298K.

### 4.3.3. New insights and constraints from our salty icy surfaces

Differences between the spectra of our analogues and previous measurements are observed and can be understood as the result of having more realistic analogues of the surface than the previous studies. Our analogues comprise neither pure water ice, nor well-crystallized pure hydrated forms of salts. They consist of intimate mixtures of the two in controlled and repeatable conditions of concentrations and crystallisation.

The absorption bands of water observed in this work result from combinations of fundamentals vibrational modes of the $H_2O$ molecule. In our analogues, the absorption features of water are distorted because of the presence of $H_2O$ both as crystalline ice and in the crystal lattice of the salt hydrates. The hydrogen bonds lengths and strengths between $H_2O$ molecules within the hydrate differ depending on the structure of the hydrate crystal lattice, which finally affects the vibrational modes of the $H_2O$ molecule.

The laboratory studies mentioned previously share several common points. Pure hydrates have been measured, and temperature effects have been noticed. The sharpness of the absorption bands within the 1.5- and 2.0-µm complexes is enhanced at low temperature. Mixtures of the salts with water have already been attempted, but either with saturated concentrations or with chemistries more complex than binary mixtures of salt and $H_2O$, therefore adding considerable complexity to the samples because of a dependence on pH for hydrates formation from solutions of three or more end-members (Johnson et al., 2019).

The global oceans predicted for Europa, and Enceladus, probably contain different salts, which justify the mixing of different solutions to mimic the surface spectra of icy moons. However, working with simple systems allow us to better identify and analyse the signatures of the different compounds.

The effect of grain size on the scattering of light by surfaces has been extensively studied. Still, in the previous studies, the influence of grain size was either established by modelling or roughly estimated. The flash-freezing procedures used in previous studies generally produced slab-like analogues (compact ice cubes), but due to the fast cooling rate, ended up showing absorption features similar to what we observe with sublimated granular salty ice particles.

The procedures for producing our analogues, and the corresponding spectral measurements, reveal that the flash-freezing procedure used in many previous studies is not the one and only way to mimic the surfaces of icy moons. Compact slab produced by the slow cooling of salty solutions are also realistic spectral analogues for the surface of icy moons (Section 4.4). Our study also points out the importance of salt concentration, which controls the shape and intensity of the absorption features of hydrates.

## 4.4. Implications for the surfaces of icy moons

Sublimation is a slow process at the surface of Europa (due to the low temperature of 70 to 135 K, (Ashkenazy, 2019; Moore and Hussmann, 2009)), expected to affect metres of water ice within several millions years. The temperature variation at the surface of Europa generates a strong latitudinal gradient of sublimation rates from the poles toward the equator (Hobley et al., 2018). It is expected that at low latitude, the sublimation process dominates over the space weathering processes, which justifies performing sublimation experiments of relevant analogues.

The relatively flat spectra associated with icy moons surfaces (section 4.4, Figure 16 and references in the caption; Ganymede: C and I; Europa: F, G and H) are best fitted by association of water ice and salts. Considerable effort has been made to mimic icy moon spectra using linear mixing models of reflectance (Brown and Hand, 2013; Dalton et al., 2012; Ligier et al., 2016, 2019; Shirley et al., 2010, 2016; Spencer et al., 2006) and propose mixtures of salts and water ice. They also concluded that hydrated compounds were relevant for icy moons surfaces.

Our compact slabs analogues show reflectance spectra with similar features to those icy moons' surfaces, as shown in Figure 16. The spectra of $MgCl_2 - H_2O$ slabs (Figure 16, E and J) and the observation of the trailing side of Europa (Figure 16, F, G, and H) are similar in the VNIR up to 1.9 µm.

Nevertheless, the absorption feature roughly at 2.01 µm in laboratory spectra (Figure 16, E and J) is not present in the observations (Figure 16, F, G and H). In ground-based data, an absorption at this wavelength can be related to a non-optimal telluric absorption (as for the $CO_2$ in Earth atmosphere). $MgSO_4$ – $H_2O$ slabs reproduce better the trailing side of Europa with lower salt concentrations (Figure 16, D and B). The case of the $Na_2SO_4$ – $H_2O$ slabs is relevant, particularly at higher concentrations (Figure 16, A: 3 wt%; K: 15 wt%), because their spectra are considerably flattened within the 1.5 – 1.8 µm and the 2.0 – 2.4 µm range.

One should notice that the least icy areas of Ganymede observed by (Ligier et al., 2019) have a stronger 1.65 µm than the non-icy area from Europa (Figure 16, C, I compared to Figure 16, H). The band at 1.65 µm indicates crystalline water ice but is also temperature dependant (Grundy and Schmitt, 1998) and would be stronger at lower temperature. This surface-feature is not as pronounced within our analogues because of (*i*) the high amount of hydrate and (*ii*) sublimation of excess water. Other observations of Europa and Ganymede have been performed by (Ligier et al., 2016, 2019) showing that the presence of hydrated salt is necessary to fit the absorption features of Ganymede and that the 2.07 µm absorption feature attributed to epsomite (Brown and Hand, 2013) would be reproduced by $MgCl_2$ hydrates. From our spectral analysis, the 2.07 µm feature has been observed in $MgSO_4$ – $H_2O$ analogues, and particularly with the 30 wt% slab. In linear modelling, the 2,07 µm "bump" can be reproduced as a consequence of the presence of a salty species rather than being an actual absorption band specific to a salt. It comes from an increase of the reflectance due to a salt, followed by a plateau (or a decrease) around 2,06 – 2,07 µm. The spectrum is, after 2,07 µm, dominated by water ice and therefore increases. This "bump" is also present in $MgCl_2$ analogues and is of main interest for linear reflectance modelling.

(Fischer et al., 2015) modelled the surface of Europa to propose components for its non-icy material. *Bloedite* ($Na_2Mg(SO_4)_2 \cdot 4H_2O$) *mirabilite* ($Na_2SO_4 \cdot 10H_2O$) and *hexahydrite* ($MgSO_4 \cdot 6H_2O$) provided good fits.

(McCord et al., 2010) reported reprocessed Galileo NIMS images of Europa to resolve detailed absorption features. Their study aimed at enhancing spectral features to identify the composition of the non-icy end-members. They reported broadening of the absorption complexes at 1.5 and 2.0 µm, a flattened region in between these two bands as well as a shift to shorter wavelengths of these bands compared to water ice-rich spectra. They also reported specific absorption features at 1.344 µm and a small feature in the 1.70 – 1.85 µm range. They explained that these features have been observed in pure sulphate ($SO_4^{2-}$) hydrates such as in (Carlson et al., 2005; Dalton, 2003; McCord et al., 2002) but we also observed the small feature in the 1.70 – 1.85 µm range within our analogues of $MgCl_2$ – $H_2O$ granular ices.

The salty slabs and granular ices we used to match the chemistries apparently required to reproduce the icy moon surfaces. No definitive match with the surface of icy moons can be given with our analogues, nevertheless it provides additional constraints on salt concentration as well as on the importance of production procedure and salt concentration to generate absorption features identified from icy moons surfaces. One future application will be the linear mixing model of reflectance taking as input parameters our analogues of intimate mixtures.

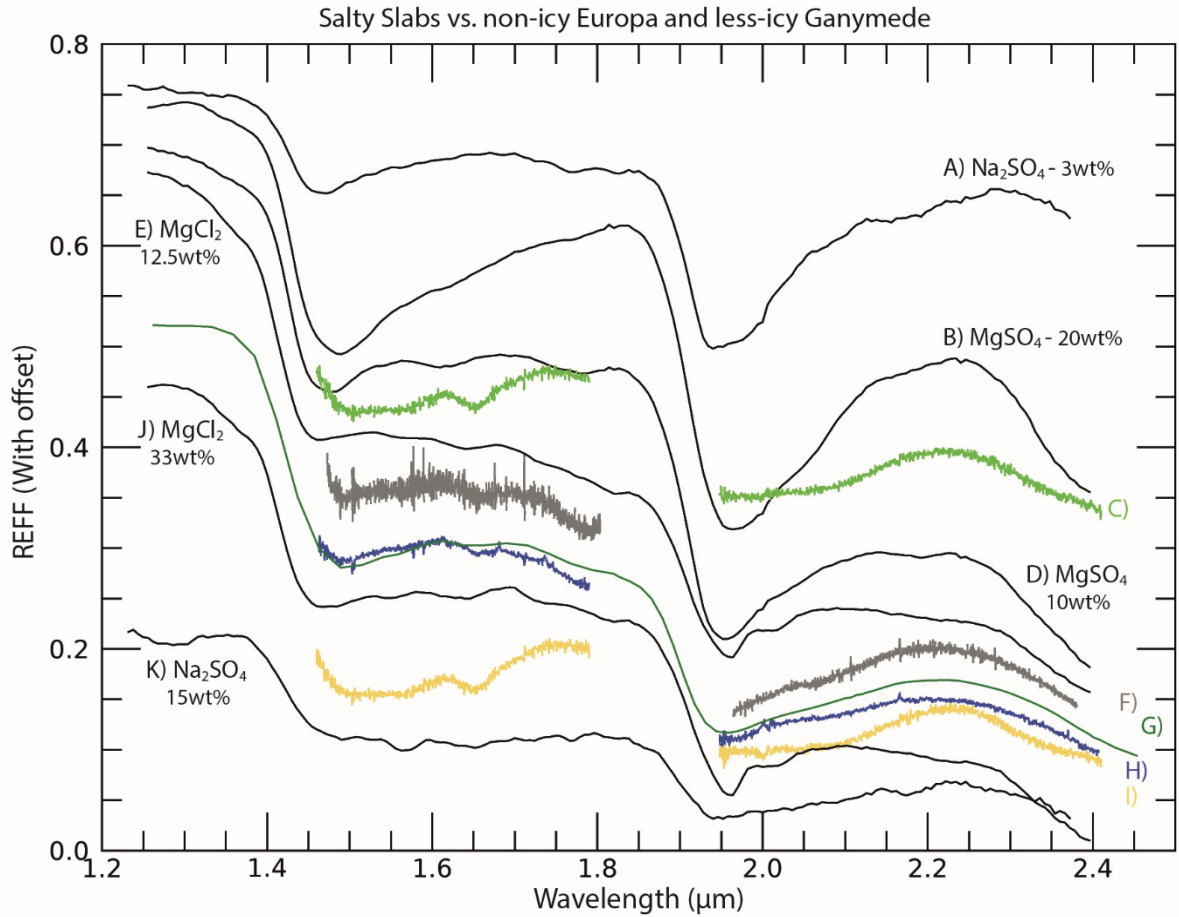

*Figure 16: reflectance spectra of salty slabs analogues compared to non-icy spectra of Europa and Ganymede. The black spectra are salty slabs of: **A** $Na_2SO_4$ – 3wt%, **B** $MgSO_4$ – 20wt% (offset +0.06), **D** $MgSO_4$ – 10wt% (offset +0.18), **E** $MgCl_2$ – 12.5 wt% (offset +0.07), **J** $MgCl_2$ – 33wt% (offset -0.2) and **K** $Na_2SO_4$ – 15wt% (offset -0.5). The coloured spectra are observation of the surface; **C** (Ligier et al., 2019) Ganymede, 300W – 30S Old Crater, Borsippa Sulcus (offset +0.20), **F** (Brown and Hand, 2013) Europa, low latitude dark trailing side, **G** (Carlson et al., 2009) Europa trailing side from the G1 ENHILAT 01a observation, **H** (Ligier et al., 2016) Europa, 260W – 10N, non-Icy area, **I** (Ligier et al., 2019) Ganymede, 140W – 20N, South Galileo Regio (offset -0.055). In ground-based observations, the region from 1.8 to 1.95 µm is not measured due to telluric absorptions.*

From our results, the initial salt concentration and the production method constrain the spectral behaviour. The results of this study form an ideal basis for tests of reflectance mixing models and direct interpretation of the ground-based observations. On the longer term, the results presented in this paper are of interest for the upcoming JUICE and Europa Clipper missions, to compare with the data that will be provided by their respective spectral imagers (MAJIS for JUICE, and MISE on Clipper).

# 5. Conclusions and perspectives

In this paper, we present a systematic study of the NIR spectra of four different salts in water ice at varying concentrations. It complements previous studies of salt hydrates and flash-frozen saturated brines cited in section 4.3. The analogues produced have a good reproducibility using tried and tested protocols and are well characterised (Jost et al., 2016; Poch et al., 2018, 2016; Pommerol et al., 2019; Yoldi et al., 2015).

The sublimation experiments revealed that the amount of hydrates formed from an initial solution is dependent on the production procedure. Slabs and granular ices produce different amounts of hydrates from the same salty liquid solution, because of their different cooling rates.

The spectra produced during this study provide new knowledge on binary mixtures of salts and water ice. The number of hydrates, as well as their level of hydration, influence the strength of the absorption features. A common trend such as systematically higher values of reflectance for granular ices compared to slabs has not been observed. The role of the salt concentration was described and also gave constraints and different trends regarding the analogue (granular vs slab). The mixing mode of salt and water has been further developed and allows identification and discrimination of hydration states regarding the analogue produced. The role of the grain size in our analogues is difficult to isolate.

Despite the spectral distortion caused by the intimate mixture of salts and water, our study shows that the spectral identification of hydrates is possible and accurate. The comparison with previous work on hydrates using Raman and VNIR spectroscopy, and the information provided by phase diagrams of salt – $H_2O$, allow a good understanding of the chemical composition of our analogues. Given the production procedure, some specific behaviour has been identified:

- The amount of hydrates within granular ices is positively correlated and dependant on the initial salt concentration.
- The most efficient procedure to maximise the number of hydrates formed is the slab production.

- The increase in concentration produced during the flash freezing procedure led the $MgCl_2$-$H_2O$ system to produce hydrates with lower hydration states that are identifiable after sublimation.

A comparison of our analogues with few ground-based and orbital observations of icy moons surfaces shows the relevance of doing intimate mixture in controlled conditions (salt concentrations, kinetic of production, grain size). It generates strong distortion of the 1.5- and 2.0-µm NIR absorption bands of water ice. Of all compositions tested so far, the best candidates to match Europa's and Ganymede's surface are slabs of $MgCl_2$, $MgSO_4$ and $Na_2SO_4$ with 15 or less wt% of salts which give new constraints compared to the saturated samples produced in previous laboratory spectral studies.

This new set of well-characterised data enriches the current dataset of icy laboratory analogues. It provides new sets of samples relevant for further icy moons experiments, such as electron irradiation of alkali halides mimicking Galilean's moon surface conditions. The database provided by our experiments constitutes a new playground for reflectance mixing models.

This work could also be extended in the laboratory, by making other types of salty analogues, performing ternary mixtures (two salts and water). Manual mixing of salts and water ices could make relevant analogues to icy moons surfaces. Other types of ices, such as $CO_2$ ices could be mixed with salty water ices, to provide more accurate analogues for the surface of Ganymede.

## Acknowledgements

The authors thank Michael E. Brown, Robert W. Carlson and Nicolas Ligier for providing their data. The team from the University of Bern is supported by the Swiss National Science Foundation, in particular through the NCCR PlanetS.

## References

Anderson, J.D., Jacobson, R.A., McElrath, T.P., Moore, W.B., Schubert, G., Thomas, P.C., 2001. Shape, mean radius, gravity field, and interior structure of Callisto. Icarus 153, 157–161. https://doi.org/10.1006/icar.2001.6664

Anderson, J.D., Schubert, G., Jacobson, R.A., Lau, E.L., Moore, W.B., Sjogren, W.L., 1998. Europa's Differentiated Internal Structure: Inferences from Four Galileo Encounters. Science (80-. ). 281, 2019–2022. https://doi.org/10.1126/science.281.5385.2019

Ashkenazy, Y., 2019. The surface temperature of Europa. Heliyon 5, e01908. https://doi.org/10.1016/j.heliyon.2019.e01908

Baumgartner, M., Bakker, R.J., 2010. Raman spectra of ice and salt hydrates in synthetic fluid inclusions. Chem. Geol. 275, 58–66. https://doi.org/10.1016/j.chemgeo.2010.04.014

Bierhaus, E.B., Zahnle, K., Chapman, C.R., 2009. Europa's Crater Distributions and Surface Ages, in: Europa. University of Arizona Press, pp. 161–180. https://doi.org/10.2307/j.ctt1xp3wdw.13

Brown, M.E., Hand, K.P., 2013. Salts and Radiation Products on the Surface of Europa. Astron. J. 145, 110. https://doi.org/10.1088/0004-6256/145/4/110

Carlson, R., Calvin, W., Dalton, J., 2009. Europa's surface composition, in: Europa. University of Arizona Press, pp. 283–327.

Carlson, R., Smythe, W., Baines, K., Barbinis, E., Becker, K., Burns, R., Calcutt, S., Calvin, W., Clark, R., Danielson, G., Davies, A., Drossart, P., Encrenaz, T., Fanale, F., Granahan, J., Hansen, G., Herrera, P., Hibbitts, C., Hui, J., Irwin, P., Johnson, T., Kamp, L., Kieffer, H., Leader, F., Lellouch, E., Lopes-Gautier, R., Matson, D., McCord, T., Mehlman, R., Ocampo, A., Orton, G., Roos-Serote, M., Segura, M., Shirley, J., Soderblom, L., Stevenson, A., Taylor, F., Torson, J., Weir, A., Weissman, P., 1996. Near-Infrared Spectroscopy and Spectral Mapping of Jupiter and the Galilean Satellites: Results from Galileo's Initial Orbit. Science (80-. ). 274, 385–388. https://doi.org/10.1126/science.274.5286.385

Carlson, R.W., 1999. Sulfuric Acid on Europa and the Radiolytic Sulfur Cycle. Science (80-. ). 286, 97–99. https://doi.org/10.1126/science.286.5437.97

Carlson, R.W., Anderson, M.S., Mehlman, R., Johnson, R.E., 2005. Distribution of hydrate on Europa: Further evidence for sulfuric acid hydrate. Icarus 177, 461–471. https://doi.org/10.1016/j.icarus.2005.03.026

Carlson, R.W., Weissman, P.R., Smythe, W.D., Mahoney, J.C., 1992. Near-Infrared Mapping Spectrometer experiment on Galileo. Space Sci. Rev. 60. https://doi.org/10.1007/BF00216865

Chyba, C.F., Hand, K.P., 2005. Astrobiology: The Study of the Living Universe. Annu. Rev. Astron. Astrophys. 43, 31–74. https://doi.org/10.1146/annurev.astro.43.051804.102202

Chyba, C.F., Phillips, C.B., 2001. Possible ecosystems and the search for life on Europa. Proc. Natl. Acad. Sci. U. S. A. https://doi.org/10.1073/pnas.98.3.801

Collins, G., Johnson, T., 2014. Ganymede and Callisto, in: Encyclopedia of the Solar System. Elsevier, pp. 449–466. https://doi.org/10.1016/B978-012088589-3/50028-1

Collins, G., Nimmo, F., 2009. Chaotic Terrain on Europa, in: Europa. University of Arizona Press, pp. 259–282. https://doi.org/10.2307/j.ctt1xp3wdw.17

Crowley, J.K., 1991. Visible and Near-Infrared (0.4-2.5 Mu-M) Reflectance Spectra of Playa Evaporite Minerals. J. Geophys. Res. Earth 96, 16231–16240. https://doi.org/10.1029/91jb01714

Dalton, J.B., 2007. Linear mixture modeling of Europa's non-ice material based on cryogenic laboratory spectroscopy. Geophys. Res. Lett. 34, 2–5. https://doi.org/10.1029/2007GL031497

Dalton, J.B., 2003. Spectral behavior of hydrated sulfate salts: implications for Europa mission spectrometer design. Astrobiology 3, 771–784. https://doi.org/10.1089/153110703322736097

Dalton, J.B., Cassidy, T., Paranicas, C., Shirley, J.H., Prockter, L.M., Kamp, L.W., 2013. Exogenic controls on sulfuric acid hydrate production at the surface of Europa. Planet. Space Sci. 77, 45–63. https://doi.org/10.1016/j.pss.2012.05.013

Dalton, J.B., Cruikshank, D.P., Stephan, K., McCord, T.B., Coustenis, A., Carlson, R.W., Coradini, A., 2010. Chemical composition of icy satellite surfaces, Space Science Reviews. https://doi.org/10.1007/s11214-010-9665-8

Dalton, J.B., Pitman, K.M., 2012. Low temperature optical constants of some hydrated sulfates relevant to planetary surfaces. J. Geophys. Res. E Planets 117, 1–15. https://doi.org/10.1029/2011JE004036

Dalton, J.B., Prieto-Ballesteros, O., Kargel, J.S., Jamieson, C.S., Jolivet, J., Quinn, R., 2005. Spectral comparison of heavily hydrated salts with disrupted terrains on Europa. Icarus 177, 472–490. https://doi.org/10.1016/j.icarus.2005.02.023

Dalton, J.B., Shirley, J.H., Kamp, L.W., 2012. Europa's icy bright plains and dark linea: Exogenic and endogenic contributions to composition and surface properties. J. Geophys. Res. 117. https://doi.org/10.1029/2011JE003909

Davis, B.L., Chevrier, V.F., Altheide, T.S., Francis, A., 2009. Reflectance Spectra of Low-Temperature Chloride and Perchlorate Hydrates and Their Relevance to the Martian Surface. New Martian Chem. Work. 3–5.

Fanale, F., Banerdt, W., Elson, L., Stjp, T.J.-, 1982, U., 1982. Io's surface-Its phase composition and influence on Io's atmosphere and Jupiter's magnetosphere, in: Satellites of Jupiter.

Feng, H., Xu, Y., Yang, T., 2018. Study on Leidenfrost effect of cryoprotectant droplets on liquid nitrogen with IR imaging technology and non-isothermal crystallization kinetics model. Int. J. Heat Mass Transf. 127, 413–421. https://doi.org/10.1016/j.ijheatmasstransfer.2018.08.001

Fischer, P.D., Brown, M.E., Hand, K.P., 2015. SPATIALLY RESOLVED SPECTROSCOPY OF EUROPA: THE DISTINCT SPECTRUM OF LARGE-SCALE CHAOS. Astron. J. 150, 164. https://doi.org/10.1088/0004-6256/150/5/164

Fox-Powell, M.G., Cousins, C.R., 2021. Partitioning of Crystalline and Amorphous Phases During Freezing of Simulated Enceladus Ocean Fluids. J. Geophys. Res. Planets 126, 1–16. https://doi.org/10.1029/2020JE006628

Gendrin, A., Mangold, N., Bibring, J.P., Langevin, Y., Gondet, B., Poulet, F., Bonello, G., Quantin, C., Mustard, J., Arvidson, R., LeMouélic, S., 2005. Sulfates in Martian layered terrains: The OMEGA/Mars express view. Science (80-. ). 307, 1587–1591. https://doi.org/10.1126/science.1109087

Grundy, W.M., Schmitt, B., 1998. The temperature-dependent near-infrared absorption spectrum of hexagonal H2O ice. J. Geophys. Res. E Planets 103, 25809–25822. https://doi.org/10.1029/98JE00738

Hamilton, A., Menzies, R.I., 2010. Raman spectra of mirabilite, Na2So4·10H2O and the rediscovered metastable heptahydrate, Na2So4·7H2o. J. Raman Spectrosc. 41, 1014–1020. https://doi.org/10.1002/jrs.2547

Hand, K.P., Carlson, R.W., 2015. Europa's surface color suggests an ocean rich with sodium chloride.

Geophys. Res. Lett. 42, 3174–3178. https://doi.org/10.1002/2015GL063559

Hand, K.P., Chyba, C.F., Priscu, J.C., Carlson, R.W., Nealson, K.H., 2009. Astrobiology and the Potential for Life on Europa, in: Pappalardo, R.T., McKinnon, W.B., Khurana, K.K. (Eds.), Europa. Univ. Ariz. Press, Tucson.

Hanley, J., Dalton, J.B., Chevrier, V.F., Jamieson, C.S., Barrows, R.S., 2014. Reflectance spectra of hydrated chlorine salts: The effect of temperature with implications for Europa 2370–2377. https://doi.org/10.1002/2013JE004565.Received

Head, J.W., Pappalardo, R.T., 1999. Brine mobilization during lithospheric heating on Europa: Implications for formation of chaos terrain, lenticula texture, and color variations. J. Geophys. Res. E Planets 104, 27143–27155. https://doi.org/10.1029/1999JE001062

Hobley, D.E.J., Moore, J.M., Howard, A.D., Umurhan, O.M., 2018. Formation of metre-scale bladed roughness on Europa's surface by ablation of ice. Nat. Geosci. 11, 901–904. https://doi.org/10.1038/s41561-018-0235-0

Hogenboom, D.L., Kargel, J.S., Ganasan, J.P., Lee, L., 1995. Magnesium Sulfate-Water to 400 MPa Using a Novel Piezometer: Densities, Phase Equilibria, and Planetological Implications. Icarus 115, 258–277. https://doi.org/10.1006/icar.1995.1096

Hudait, A., Molinero, V., 2014. Ice crystallization in ultrafine water-salt aerosols: Nucleation, ice-solution equilibrium, and internal structure. J. Am. Chem. Soc. 136, 8081–8093. https://doi.org/10.1021/ja503311r

Jia, X., Kivelson, M.G., Khurana, K.K., Kurth, W.S., 2018. Evidence of a plume on Europa from Galileo magnetic and plasma wave signatures. Nat. Astron. 2, 459–464. https://doi.org/10.1038/s41550-018-0450-z

Johnson, P. V., Hodyss, R., Vu, T.H., Choukroun, M., 2019. Insights into Europa's ocean composition derived from its surface expression. Icarus 321, 857–865. https://doi.org/10.1016/j.icarus.2018.12.009

Jost, B., Gundlach, B., Pommerol, A., Oesert, J., Gorb, S.N., Blum, J., Thomas, N., 2013. Micrometer-sized ice particles for planetary-science experiments - II. Bidirectional reflectance. Icarus 225, 352–366. https://doi.org/10.1016/j.icarus.2013.04.007

Jost, B., Pommerol, A., Poch, O., Brouet, Y., Fornasier, S., Carrasco, N., Szopa, C., Thomas, N., 2017. Bidirectional reflectance of laboratory cometary analogues to interpret the spectrophotometric properties of the nucleus of comet 67P/Churyumov-Gerasimenko. Planet. Space Sci. 148, 1–11. https://doi.org/10.1016/j.pss.2017.09.009

Jost, B., Pommerol, A., Poch, O., Gundlach, B., Leboeuf, M., Dadras, M., Blum, J., Thomas, N., 2016. Experimental characterization of the opposition surge in fine-grained water-ice and high albedo ice analogs. Icarus 264, 109–131. https://doi.org/10.1016/j.icarus.2015.09.020

Kargel, J.S., 1991. Brine volcanism and the interior structures of asteroids and icy satellites. Icarus 94, 368–390. https://doi.org/10.1016/0019-1035(91)90235-L

Kargel, J.S., Kaye, J.Z., Head, J.W., Marion, G.M., Sassen, R., Crowley, J.K., Ballesteros, O.P., Grant, S.A., Hogenboom, D.L., 2000. Europa's Crust and Ocean: Origin, Composition, and the Prospects for Life. Icarus 148, 226–265. https://doi.org/10.1006/icar.2000.6471

Khurana, K.K., Kivelson, M.G., Stevenson, D.J., Schubert, G., Russell, C.T., Walker, R.J., Polanskey, C., 1998. Induced magnetic fields as evidence for subsurface oceans in Europa and Callisto. Nature 395, 777–780. https://doi.org/10.1038/27394

Kivelson, M.G., 2000. Galileo Magnetometer Measurements: A Stronger Case for a Subsurface Ocean at Europa. Science (80-. ). 289, 1340–1343. https://doi.org/10.1126/science.289.5483.1340

Kivelson, M.G., Khurana, K.K., Stevenson, D.J., Bennett, L., Joy, S., Russell, C.T., Walker, R.J., Zimmer, C., Polanskey, C., 1999. Europa and Callisto: Induced or intrinsic fields in a periodically varying plasma environment. J. Geophys. Res. Sp. Phys. 104, 4609–4625. https://doi.org/10.1029/1998JA900095

Kivelson, M.G., Khurana, K.K., Volwerk, M., 2002. The permanent and inductive magnetic moments of Ganymede. Icarus 157, 507–522. https://doi.org/10.1006/icar.2002.6834

Le Gall, A., Leyrat, C., Janssen, M.A., Choblet, G., Tobie, G., Bourgeois, O., Lucas, A., Sotin, C., Howett, C., Kirk, R., Lorenz, R.D., West, R.D., Stolzenbach, A., Massé, M., Hayes, A.H., Bonnefoy, L., Veyssière, G., Paganelli, F., 2017. Thermally anomalous features in the subsurface of Enceladus's south polar terrain. Nat. Astron. 1, 0063. https://doi.org/10.1038/s41550-017-0063

Li, D., Zeng, D., Yin, X., Han, H., Guo, L., Yao, Y., 2016. Phase diagrams and thermochemical modeling of salt lake brine systems. II. NaCl+H2O, KCl+H2O, MgCl2+H2O and CaCl2+H2O systems. Calphad Comput. Coupling Phase Diagrams Thermochem. 53, 78–89. https://doi.org/10.1016/j.calphad.2016.03.007

Ligier, N., Carter, J., Poulet, F., Langevin, Y., Shirley, J.., 2017. High spatial and spectral resolution near infrared mapping of Ganymede and Callisto with ESO/VLT/SINFONI 6–7.

Ligier, N., Paranicas, C., Carter, J., Poulet, F., Calvin, W.M., Nordheim, T.A., Snodgrass, C., Ferellec, L., 2019. Surface composition and properties of Ganymede: Updates from ground-based observations with the near-infrared imaging spectrometer SINFONI/VLT/ESO. Icarus 333, 496–515. https://doi.org/10.1016/j.icarus.2019.06.013

Ligier, N., Poulet, F., Carter, J., Brunetto, R., Gourgeot, F., 2016. VLT/SINFONI Observations of Europa: New insights into the surface composition. Astron. J. 151, 163. https://doi.org/10.3847/0004-6256/151/6/163

Ludl, A.-A., Bove, L.E., Li, J., Morand, M., Klotz, S., 2017. Quenching device for electrolytic aqueous solutions. Eur. Phys. J. Spec. Top. 226, 1051–1063. https://doi.org/10.1140/epjst/e2016-60244-8

Mayer, E., 1985. New method for vitrifying water and other liquids by rapid cooling of their aerosols. J. Appl. Phys. 58, 663–667. https://doi.org/10.1063/1.336179

McCarthy, C., Cooper, R.F., Kirby, S.H., Rieck, K.D., Stern, L.A., 2007. Solidification and microstructures of binary ice-I/hydrate eutectic aggregates. Am. Mineral. 92, 1550–1560. https://doi.org/10.2138/am.2007.2435

McCord, T.B., Hansen, G.B., Combe, J.P., Hayne, P., 2010. Hydrated minerals on Europa's surface: An improved look from the Galileo NIMS investigation. Icarus 209, 639–650. https://doi.org/10.1016/j.icarus.2010.05.026

McCord, T.B., Hansen, G.B., Fanale, F.P., Carlson, R.W., Maison, D.L., Johnson, T. V, Smythe, W.D., Crowley, J.K., Martin, P.D., Ocampo, A., Hibbitts, C.A., Granahan, J.C., 1998. Salts on Europa ' s Surface Detected by Galileo ' s Near Infrared Mapping Spectrometer 280. https://doi.org/10.1126/science.280.5367.1242

McCord, T.B., Hansen, G.B., Matson, D.L., Johnson, T. V, Crowley, J.K., Fanale, F.P., Carlson, R.W., Smythe, W.D., Martin, P.D., Hibbitts, C.A., Granahan, J.C., Ocampo, A., 1999. Hydrated salt minerals on Europa's Surface from the Galileo near-infrared mapping spectrometer (NIMS) investigation - paper a. J. Geophys. Res. 104, 11827–11851.

https://doi.org/10.1029/1999JE900005

McCord, T.B., Orlando, T.M., Teeter, G., Hansen, G.B., Sieger, M.T., Petrik, N.G., Van Keulen, L., 2001. Thermal and radiation stability of the hydrated salt minerals epsomite, mirabilite, and natron under Europa environmental conditions. J. Geophys. Res. Planets 106, 3311–3319. https://doi.org/10.1029/2000JE001282

McCord, T.B.T., Teeter, G., Hansen, G.B.G., Sieger, M.T., Orlando, T.M., 2002. Brines exposed to Europa surface conditions. J. Geophys. … 107, 1–6. https://doi.org/10.1029/2000JE001453

Moore, M.H., Hudson, R.L., Carlson, R.W., 2007. The radiolysis of SO2 and H2S in water ice: Implications for the icy jovian satellites. Icarus 189, 409–423. https://doi.org/10.1016/j.icarus.2007.01.018

Moore, W.B., Hussmann, H., 2009. Thermal Evolution of Europa's Silicate Interior, in: Europa. University of Arizona Press.

Nagel, K., Breuer, D., Spohn, T., 2004. A model for the interior structure, evolution, and differentation of Callisto. Icarus 169, 402–412. https://doi.org/10.1016/j.icarus.2003.12.019

Orlando, T.M., McCord, T.B., Grieves, G.A., 2005. The chemical nature of Europa surface material and the relation to a subsurface ocean. Icarus 177, 528–533. https://doi.org/10.1016/j.icarus.2005.05.009

Pappalardo, R.T., Belton, M.J.S., Breneman, H.H., Carr, M.H., Chapman, C.R., Collins, G.C., Denk, T., Fagents, S., Geissler, P.E., Giese, B., Greeley, R., Greenberg, R., Head, J.W., Helfenstein, P., Hoppa, G., Kadel, S.D., Klaasen, K.P., Klemaszewski, J.E., Magee, K., McEwen, a. S., Moore, J.M., Moore, W.B., Neukum, G., Phillips, C.B., Prockter, L.M., Schubert, G., Senske, D. a., Sullivan, R.J., Tufts, B.R., Turtle, E.P., Wagner, R., Williams, K.K., 1999. Does Europa have a subsurface ocean? Evaluation of the geological evidence. J. Geophys. Res. 104, 15–55. https://doi.org/10.1029/1998JE000628

Peterson, R.C., Wang, R., 2006. Crystal molds on Mars: Melting of a possible new mineral species to create Martian chaotic terrain. Geology 34, 957. https://doi.org/10.1130/G22678A.1

Poch, O., Cerubini, R., Pommerol, A., Jost, B., Thomas, N., 2018. Polarimetry of Water Ice Particles Providing Insights on Grain Size and Degree of Sintering on Icy Planetary Surfaces. J. Geophys. Res. Planets 123, 2564–2584. https://doi.org/10.1029/2018JE005753

Poch, O., Pommerol, A., Jost, B., Carrasco, N., Szopa, C., Thomas, N., 2016. Sublimation of ice-tholins mixtures: A morphological and spectro-photometric study. Icarus 266, 288–305. https://doi.org/10.1016/j.icarus.2015.11.006

Pommerol, A., Jost, B., Poch, O., El-Maarry, M.R., Vuitel, B., Thomas, N., 2015. The SCITEAS experiment: Optical characterizations of sublimating icy planetary analogues. Planet. Space Sci. 109–110, 106–122. https://doi.org/http://dx.doi.org/10.1016/j.pss.2015.02.004

Pommerol, A., Jost, B., Poch, O., Yoldi, Z., Brouet, Y., Gracia-Berná, A., Cerubini, R., Galli, A., Wurz, P., Gundlach, B., Blum, J., Carrasco, N., Szopa, C., Thomas, N., 2019. Experimenting with Mixtures of Water Ice and Dust as Analogues for Icy Planetary Material. Space Sci. Rev. 215, 37. https://doi.org/10.1007/s11214-019-0603-0

Postberg, F., Kempf, S., Schmidt, J., Brilliantov, N., Beinsen, A., Abel, B., Buck, U., Srama, R., 2009. Sodium salts in E-ring ice grains from an ocean below the surface of Enceladus. Nature 459, 1098–1101. https://doi.org/10.1038/nature08046

Postberg, F., Schmidt, J., Hillier, J., Kempf, S., Srama, R., 2011. A salt-water reservoir as the source of a compositionally stratified plume on Enceladus. Nature 474, 620–622. https://doi.org/10.1038/nature10175

Poston, M.J., Carlson, R.W., Hand, K.P., 2017. Spectral Behavior of Irradiated Sodium Chloride Crystals Under Europa-Like Conditions. J. Geophys. Res. Planets 122, 2644–2654. https://doi.org/10.1002/2017JE005429

Prockter, L.M., Patterson, G.W., 2009. Morphology and Evolution of Europa's Ridges and Bands, in: Europa. University of Arizona Press, pp. 237–258. https://doi.org/10.2307/j.ctt1xp3wdw.16

Roth, L., Saur, J., Retherford, K.D., Strobel, D.F., Feldman, P.D., McGrath, M. a, Nimmo, F., 2014. Transient water vapor at Europa's south pole. Science (80-. ). 343, 171–174. https://doi.org/10.1126/science.1247051

Samson, I.M., Walker, R.T., 2000. Cryogenic raman spectroscopic studies in the system NaCl-CaCl2-H2O and implications for low temperature phase behavior in aqueous fluid inclusions. Can. Mineral. 38, 35–43. https://doi.org/10.2113/gscanmin.38.1.35

Schmitt, B., Quirico, E., Trotta, F., Grundy, W.M., 1998. Optical Properties of Ices From UV to Infrared 199–240. https://doi.org/10.1007/978-94-011-5252-5_9

Shi, E., Wang, A., Ling, Z., 2019. MIR, VNIR, NIR, and Raman spectra of magnesium chlorides with six hydration degrees: Implication for Mars and Europa. J. Raman Spectrosc. 51, 1589–1602. https://doi.org/10.1002/jrs.5700

Shirley, J.H., Dalton, J.B., Prockter, L.M., Kamp, L.W., 2010. Europa's ridged plains and smooth low albedo plains: Distinctive compositions and compositional gradients at the leading side-trailing side boundary. Icarus 210, 358–384. https://doi.org/10.1016/j.icarus.2010.06.018

Shirley, J.H., Jamieson, C.S., Dalton, J.B., 2016. Europa's surface composition from near-infrared observations: A comparison of results from linear mixture modeling and radiative transfer modeling. Earth Sp. Sci. 3, 326–344. https://doi.org/10.1002/2015EA000149

Sotin, C., Tobie, G., 2004. Internal structure and dynamics of the large icy satellites. Comptes Rendus Phys. 5, 769–780. https://doi.org/10.1016/j.crhy.2004.08.001

Sparks, W.B., Hand, K.P., McGrath, M.A., Bergeron, E., Cracraft, M., Deustua, S.E., 2016. Probing for Evidence of Plumes on Europa With Hst /Stis . Astrophys. J. 829, 121. https://doi.org/10.3847/0004-637x/829/2/121

Spencer, J.R., Grundy, W.M., Dumas, C., Carlson, R.W., McCord, T.B., Hansen, G.B., Terrile, R.J., 2006. The nature of Europa's dark non-ice surface material: Spatially-resolved high spectral resolution spectroscopy from the Keck telescope. Icarus 182, 202–210. https://doi.org/10.1016/j.icarus.2005.12.024

Stevenson, D.J., 1996. When Galileo met Ganymede, in: Nature. pp. 511–512. https://doi.org/10.1038/384511a0

Thomas, E.C., Hodyss, R., Vu, T.H., Johnson, P. V., Choukroun, M., 2017. Composition and Evolution of Frozen Chloride Brines under the Surface Conditions of Europa. ACS Earth Sp. Chem. 1, 14–23. https://doi.org/10.1021/acsearthspacechem.6b00003

Trumbo, S.K., Brown, M.E., Hand, K.P., 2019. Sodium Chloride on the Surface of Europa. Sci. Adv. 2–6.

Vu, T.H., Hodyss, R., Choukroun, M., Johnson, P. V., 2016. Chemistry of Frozen Sodium–Magnesium–Sulfate–Chloride Brines: Implications for Surface Expression of Europa'S Ocean Composition. Astrophys. J. 816, L26. https://doi.org/10.3847/2041-8205/816/2/l26

Waite, J.H., 2006. Cassini Ion and Neutral Mass Spectrometer: Enceladus Plume Composition and Structure. Science (80-. ). 311, 1419–1422. https://doi.org/10.1126/science.1121290

Waite, J.H., Glein, C.R., Perryman, R.S., Teolis, B.D., Magee, B.A., Miller, G., Grimes, J., Perry, M.E., Miller, K.E., Bouquet, A., Lunine, J.I., Brockwell, T., Bolton, S.J., 2017. Cassini finds molecular hydrogen in the Enceladus plume: Evidence for hydrothermal processes. Science (80-. ). 356, 155–159. https://doi.org/10.1126/science.aai8703

Wang, A., Freeman, J.J., Chou, I.-M., Jolliff, B.L., 2011. Stability of Mg-sulfates at −10°C and the rates of dehydration/rehydration processes under conditions relevant to Mars. J. Geophys. Res. 116, E12006. https://doi.org/10.1029/2011JE003818

Wang, A., Freeman, J.J., Jolliff, B.L., Chou, I.M., 2006. Sulfates on Mars: A systematic Raman spectroscopic study of hydration states of magnesium sulfates. Geochim. Cosmochim. Acta 70, 6118–6135. https://doi.org/10.1016/j.gca.2006.05.022

Yeşilbaş, M., Lee, C.C., Boily, J.-F., 2018. Ice and Cryosalt Formation in Saline Microporous Clay Gels. ACS Earth Sp. Chem. 2, 314–319. https://doi.org/10.1021/acsearthspacechem.7b00134

Yoldi, Z., Pommerol, A., Jost, B., Poch, O., Gouman, J., Thomas, N., 2015. VIS-NIR reflectance of water ice/regolith analogue mixtures and implications for the detectability of ice mixed within planetary regoliths. Geophys. Res. Lett. 42, 6205–6212. https://doi.org/10.1002/2015GL064780

Zahnle, K., Alvarellos, J.L., Dobrovolskis, A., Hamill, P., 2008. Secondary and sesquinary craters on Europa. Icarus 194, 660–674. https://doi.org/10.1016/j.icarus.2007.10.024

Zahnle, K., Schenk, P., Levison, H., Dones, L., 2003. Cratering rates in the outer Solar System. Icarus 163, 263–289. https://doi.org/10.1016/S0019-1035(03)00048-4

Zolotov, M.Y., Kargel, J.S., 2009. On the Chemical Composition of Europa's Icy Shell, Ocean, and Underlying Rocks, in: Europa. University of Arizona Press, pp. 431–458. https://doi.org/10.2307/j.ctt1xp3wdw.24

Zolotov, M.Y., Shock, E.L., 2001. Composition and stability of salts on the surface of Europa and their oceanic origin. J. Geophys. Res. Planets 106, 32815–32827. https://doi.org/10.1029/2000JE001413